\documentclass[10pt,a4paper,twoside]{article}
\usepackage{txfonts}
\usepackage{graphicx}
\usepackage{fancyhdr}
\newcommand {\sect}[1]{\section{#1} \rule[5mm]{1\headwidth}{0.5mm}}

\begin{document}

\begin{titlepage}
\vspace*{-2cm}
\begin{flushright}
  physics/0509016\\
  CARE/ELAN Document-2005-013\\
  CLIC Note 639\\
  KEK Preprint 2005-60\\
  LAL 05-94\\
  September 2, 2005
\end{flushright}
\begin{center}
{\Large
Conceptual Design of a Polarised Positron Source Based on Laser Compton
  Scattering
  \protect\\
  {\it --- A Proposal Submitted to Snowmass 2005 --- } } 

 \vspace*{5 mm} 
 Sakae Araki,
 Yasuo Higashi,
 Yousuke Honda,
 Yoshimasa Kurihara,
 Masao Kuriki,  \\
 Toshiyuki Okugi,
 Tsunehiko Omori, 
 Takashi Taniguchi,
 Nobuhiro Terunuma,
 Junji Urakawa
 \protect\\
 {\it (KEK, Ibaraki, Japan)}
 \vspace*{2 mm} \protect\\  

 X.~Artru, 
 M.~Chevallier \protect\\
 {\it (IPN, Lyon, France)}
 \vspace*{2 mm} \protect\\  

 V.~Strakhovenko \protect\\
 {\it (BINP, Novosibirsk, Russia)}
 \vspace*{5 mm} \protect\\  

 Eugene Bulyak,
 Peter Gladkikh \protect\\
 {\it (NSC KIPT, Kharkov, Ukraine)}
 \vspace*{2 mm} \protect\\  

 Klaus M\"onig \protect\\
 {\it (DESY, Zeuthen, Germany \& LAL, Orsay, France)} 
 \vspace*{2 mm} \protect\\  

 Robert Chehab,
 Alessandro Variola,
 Fabian Zomer
 \protect\\
 {\it (LAL, Orsay, France)}
 \vspace*{2 mm} \protect\\  

  Susanna Guiducci,
  Pantaleo Raimondi
 \protect\\
 {\it (INFN, Frascati, Italy)}
 \vspace*{2 mm} \protect\\  

 Frank Zimmermann\protect\\
 {\it (CERN, Geneva, Switzerland)}
  \vspace*{2 mm} \protect\\  

 Kazuyuki Sakaue,
 Tachishige Hirose,
 Masakazu Washio
\protect\\
 {\it (Waseda University, Tokyo, Japan)}
 \vspace*{2 mm} \protect\\  

 Noboru Sasao,
 Hirokazu Yokoyama
\protect\\
 {\it (Kyoto University, Kyoto, Japan)}
 \vspace*{2 mm} \protect\\  

 Masafumi Fukuda,
 Koichiro Hirano,
 Mikio Takano
\protect\\
 {\it (NIRS, Chiba, Japan)}
 \vspace*{2 mm} \protect\\  

 Tohru Takahashi,
 Hiroki Sato
\protect\\
 {\it (Hiroshima University, Hiroshima, Japan)}
 \vspace*{2 mm} \protect\\  

 Akira Tsunemi
\protect\\
 {\it (Sumitomo Heavy Industries Ltd., Tokyo, Japan)}
 \vspace*{2 mm} \protect\\  

 Jie Gao
\protect\\
 {\it (IHEP, Beijing, China)}
 \vspace*{2 mm} \protect\\

Viktor Soskov 
\protect\\
 {\it (IHEP, P.N. Lebedev Physical Institute, Russian Academy of Sciences,
Moscow)}
 \vspace*{2 mm} \protect\\

 
\end{center}
\end{titlepage}
\clearpage
\thispagestyle{empty}

\subsection*{Abstract}
We describe a scheme for producing polarised positrons
at the ILC from polarised X-rays created by Compton
scattering of a few-GeV electron beam off a ${\rm CO}_2$ or YAG laser.
This scheme is very energy effective using high finesse laser cavities in
conjunction with an electron storage ring.

\clearpage
\setcounter{page}{1}
\sect{Introduction}
\def\GeV{\ifmmode {\,\mathrm{ Ge\kern -0.1em V}}\else
                   \textrm{Ge\kern -0.1em V}\fi}%
\def\MeV{\ifmmode {\,\mathrm{ Me\kern -0.1em V}}\else
                   \textrm{Me\kern -0.1em V}\fi}%
\newcommand{\micron}{\mu\mathrm{m}}

At the ILC there exists a well motivated physics case to have not only the
electron beam, but also the positron beam polarised \cite{power}.
Up to now two ideas to produce polarised positrons have been studied in
detail. Both schemes first produce polarised photons which are then converted
into polarised positrons in a thin target.
In the first scheme the polarised positrons are produced in a helical
undulator by the high energy electron beam upstream of the interaction
point \cite{undulator}. In the second scheme the photons are produced by
Compton scattering of a 6\,GeV electron beam off a ${\rm CO}_2$ laser
\cite{posic}. 

The undulator scheme seems technically easier.
However it requires the electron linac to be fully operational before positrons
can be produced and it adds an additional energy spread of 0.15\% to the
electron beam. 
Also some important problems like the vacuum in the undulator and radiation on
the beampipe inside the undulator still need to be solved.
For the Compton scattering solution a high current electron beam
needs to be produced and a complicated high power laser scheme is needed.
However the positron polarisation can be switched easily by switching the
laser polarisation and the reachable degree of polarisation is in principle
higher than for the undulator scheme.

In this note some ideas are presented on how the time structure of the ILC
could be utilised to produce polarised positrons by Compton scattering of an
electron beam in a storage ring and a laser resonator.
R\&D for technical improvements toward an high finesse resonator 
(Gain$ > 10000$, see appendix \ref{sec:rdorsay})
would obviuosly strongly reduce the
cost and the complexity of the proposed design.

This note is meant as a basis for first discussions with accelerator experts to
understand if this scheme is possible at all.

\subsection{Basics of Compton scattering}
The kinematics and cross section of Compton scattering is governed by
the variable
\[
x=\frac{4E_b \omega_0 }{m^2c^4}\cos^2{\frac{\alpha}{2}}
 \simeq 
 0.019\left[\frac{E_b}{\GeV}\right]
\left[\frac{\mu m}{\lambda}\right],
\]
representing the scaled squared centre of mass energy of the
electron-photon system.
$E_b$ denotes the electron beam energy and $\omega_0$ the energy of
the laser photons. $\alpha$ is the crossing angle of the electron beam
and the laser. The maximum energy of the scattered photons is
given by $E_\gamma < x/(x+1) E_b$. Relevant $x$ values for polarised
positron production are $x = {\cal O}(0.01)$.

Since Compton scattering conserves parity, the total and differential
cross section cannot depend on the individual electron and laser
polarisations, but only on their product ${\cal P}_e \lambda_\gamma$.

\begin{figure}[htbp]
  \centering
  \includegraphics[width=0.48\linewidth,bb=11 7 502 466]{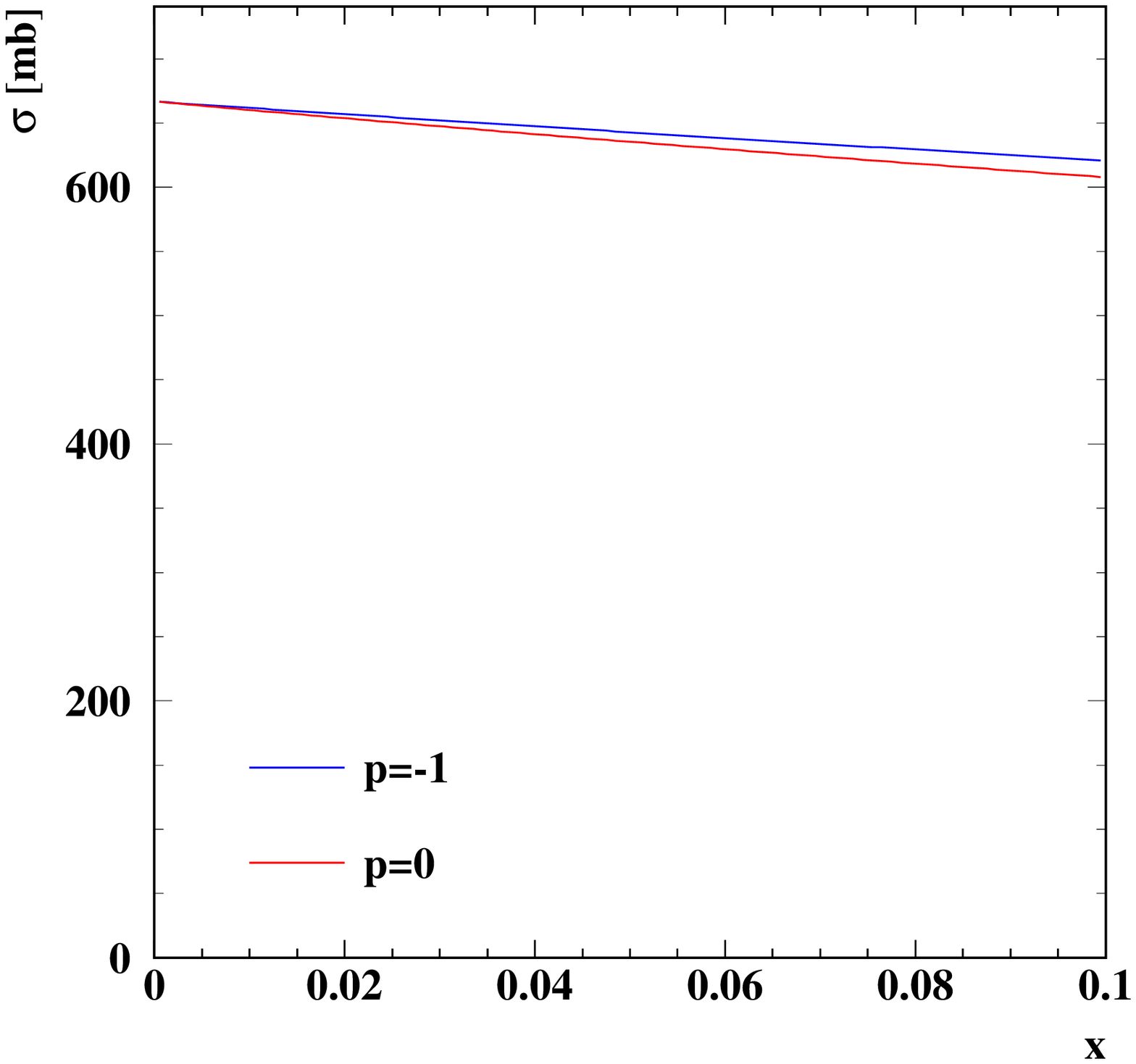}
  \includegraphics[width=0.48\linewidth,bb=8 2 507 466]{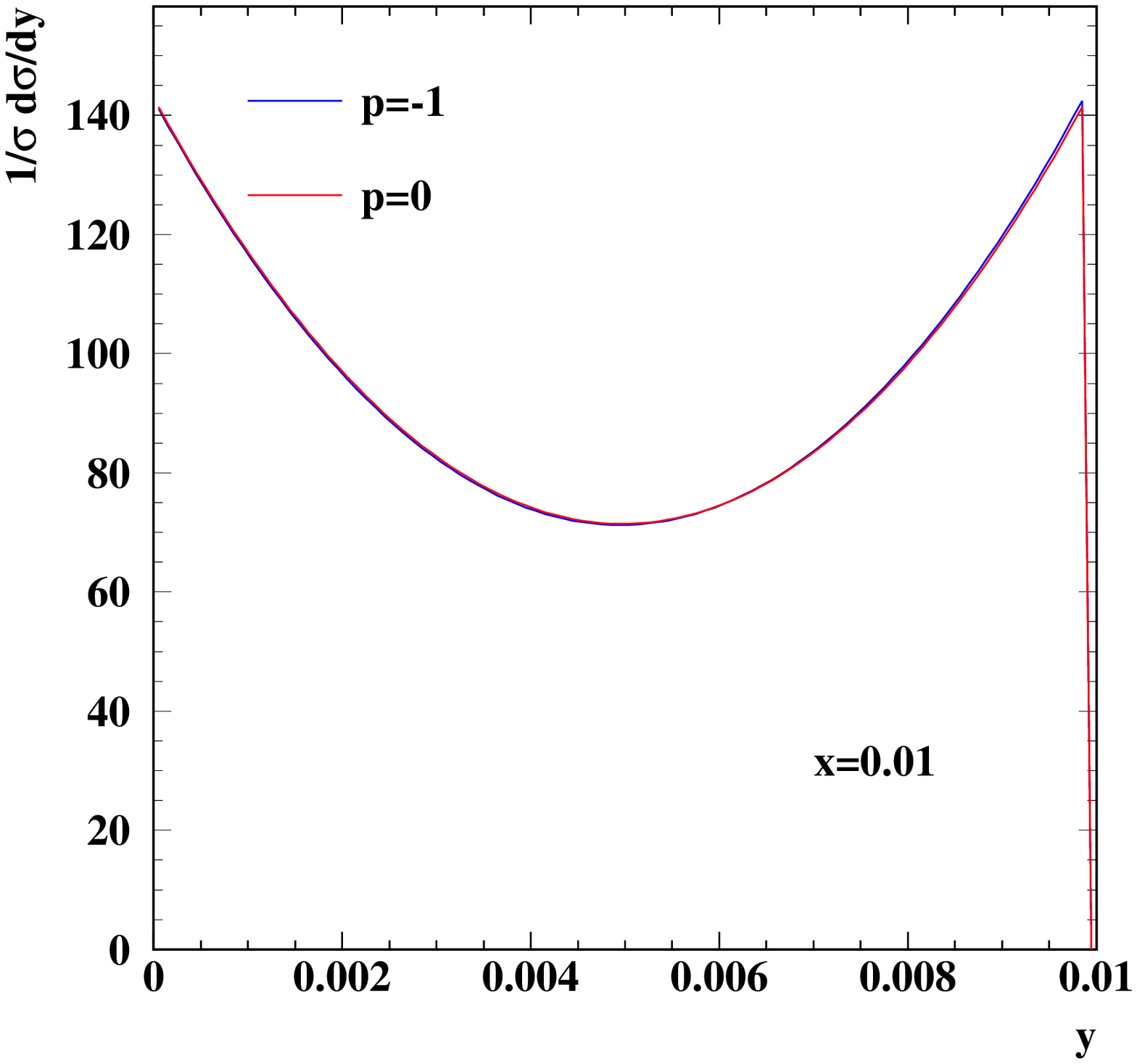}
  \caption{Total Compton cross section (left) and photon energy
    spectrum (right) for different electron/laser polarisation. P
    denotes the product of the two polarisations, y denotes the photon energy
    scaled to the beam energy.}
  \label{fig:csec}
\end{figure}

Figure \ref{fig:csec} shows the total cross section as a function of
$x$ and the photon energy spectrum for $x=0.01$ for 
${\cal P}_e \lambda_\gamma=0$ and ${\cal P}_e \lambda_\gamma=-1$. The
$x$ dependence of the cross section in the relevant range is small and
the polarisation dependence almost negligible. The photon spectrum
shows two peaks at zero and at maximum energy.

\begin{figure}[htbp]
  \centering
  \includegraphics[width=0.5\linewidth,bb=8 2 507 466]{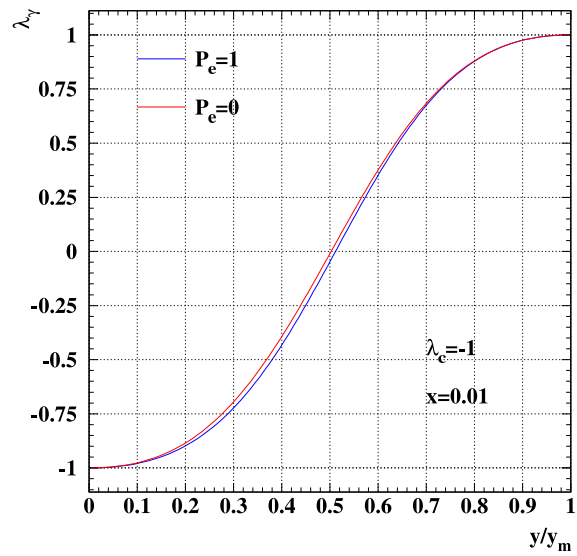}  
  \caption{Photo polarisation as a function of $x$ for 100\% laser
  polarisation and different electron polarisations}
  \label{fig:cpol}
\end{figure}

The scattered photon polarisation depends on the laser and electron
polarisation separately and is shown in figure \ref{fig:cpol} as a
function of the photon energy for ${\cal P}_e \lambda_\gamma=-1$
The dependence on the electron polarisation is very small at small $x$ so that
there is no need for electron polarisation in the storage ring. For a
highly polarised laser very high $\gamma$ polarisation can be achieved at high
energy. Since also the polarisation transfer to the positron in the pair
production works best at high energy transfer the high energy positrons will
be highly polarised.

\subsection{The laser cavity scheme}
\begin{figure}[p]
  \centering
  \includegraphics[height=0.9\textheight]{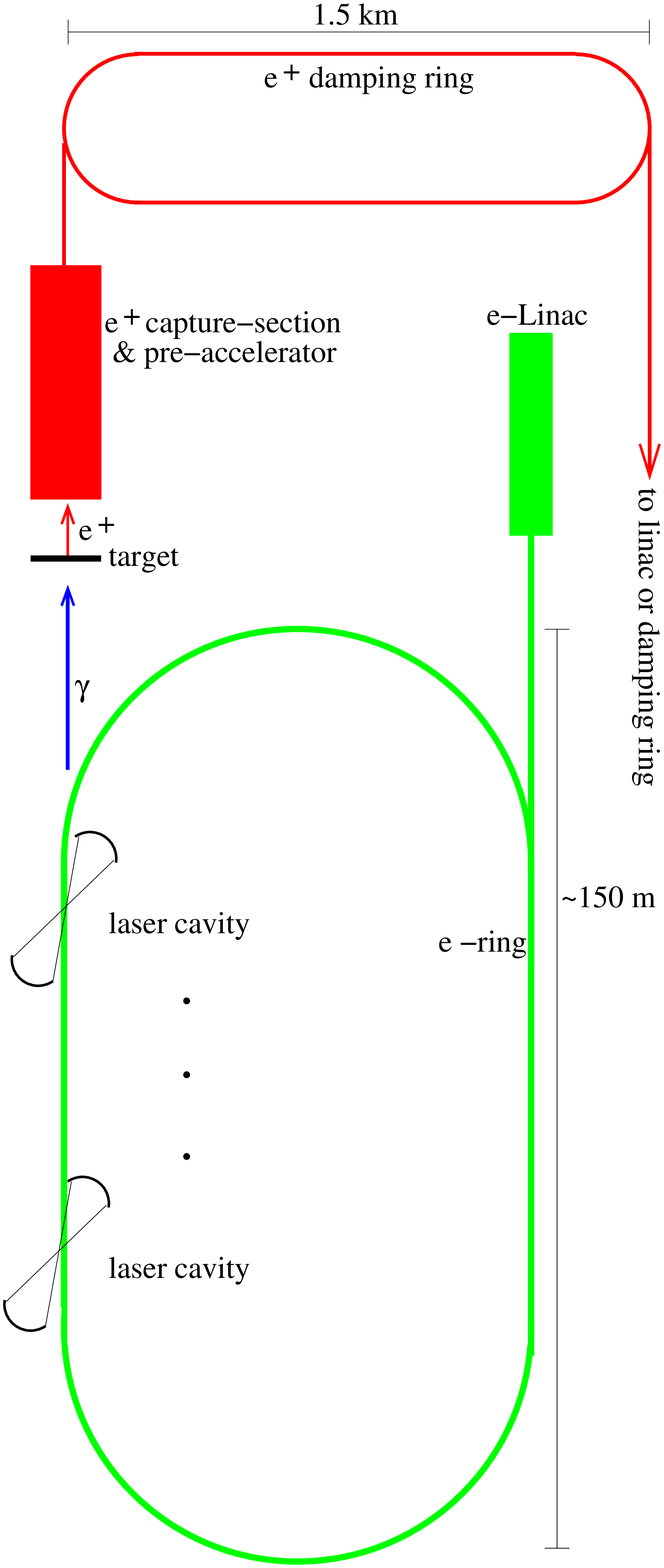}
  \caption{Principle scheme of the proposed system}
  \label{fig:scheme}
\end{figure}

In the damping ring the $e^+$ are stored with a much smaller distance than in
the main accelerator. In this note it will be assumed that a distance of 3\,ns
is possible.  This bunch spacing can also be used for the positron production.
It might thus be possible to store the electrons for the Compton scattering in
a storage ring with 1\,m bunch spacing and collide them in one or few points
with a laser cavity. The length of this storage ring is arbitrary too a large
extent. The photons are converted into positrons in a thin target and the
positrons are captured in the same way as in the other schemes. The positrons
are then accumulated in the damping ring or in a special accumulator ring.  If
it is technically possible accumulation in the damping ring saves the cost of
the extra accumulator ring. However in this case the time between two trains
needs to be shared between positron accumulation and damping.  A sketch of the
proposed scheme is shown in figure \ref{fig:scheme}.  If the full time can be
used for positron generation each bunch can receive positrons from about 20000
beam-laser interactions.

For the laser cavity two concepts will presented, wither a solid state laser,
like Nd:Yag with a wavelength of $\lambda = 1.06 \micron$ or a ${\rm CO}_2$
laser with a wavelength of $\lambda = 10.6 \micron$. For identical laser
parameters the ${\rm CO}_2$ has ten times more photons/pulse than the Nd:Yag,
however in practice the Nd:Yag laser can be focused to a significantly
smaller spot size. 
Furthermore the Nd:Yag laser technology is better suited to develop high
finesse resonators to improve the photon intensity.
Also the Compton ring energy needs to be higher in the 
${\rm CO}_2$ version to obtain the same scattered photon energy.


\clearpage
\sect{Compton Ring Design and Simulation of Compton Scattering}

\subsection{Necessary frame parameters}
Necessary frame parameters of the Compton Rings (CR) are given and
listed in Table~\ref{listpara}. CR should be capable
to emit $1.39\times 10^{10}$ gamma-quanta within the energy range
$23.2\ldots 29$\,MeV per bunch passed through the interaction
section.

\begin{table}[hbtp]
\caption{List of parameters of Compton rings\label{listpara}}
\centering 
\vspace{2pt}
\begin{tabular}{|l|r|r|}
\hline parameter &CO2 & YAG\\[1ex]
\hline
Electron energy (GeV) &  4.1 & 1.3 \\[1ex]
Electron bunch charge (nC) &  10 & 10\\[1ex]
RF frequency (MHz) & 650  & 650 \\[1ex]
Hor beam size at IP, rms ($\mu$m)&  25 & 25\\[1ex]
Ver beam size at IP, rms ($\mu$m)&  5 & 5\\[1ex]
Bunch length at IP, rms (mm)&  5 & 5\\[1ex]
\hline
Laser photon energy (eV) & 0.116 & 1.164\\[1ex]
Laser radius at IP, rms ($\mu$m ) & 25 & 5\\[1ex]
Laser pulse width, rms (mm) &  0.9 & 0.9\\[1ex]
Laser pulse power / cavity (mJ)& 210 & 592\\[1ex]
Number of laser cavities (IPs) & 30  & 30\\[1ex]
Crossing angle (degrees)& 8  & 8\\[1ex]
\hline
\end{tabular}
\end{table}

\paragraph{Remarks to the table:}
The transverse bunch dimensions at the interaction points (IPs) are
consistent with the horizontal emittance $0.5$\,nm\,rad, the
coupling $0.02$, and $\beta_{x,z}$--function values $1.25$\,m and $2.5$\,m,
respectively.
The bunch length is not directly employed in the calculations.

\subsection{Brief review of laser--electron interactions} To
estimate feasibility of Compton ring, here a brief classification of
photon--electron interaction is presented.

Electron--photon interactions mainly governed by two parameters: (i)
the density of photons (field strength in classical
electrodynamics), and (ii) the energy of each laser photon
(frequency in classical electrodynamics). We will refer to the first
parameter as the parameter of nonlinearity, to the second as the
quantumness (essentially quantum description of the process).

For the Compton ring it is expedient to present the parameter of
nonlinearity (see \cite{goldman64e,khokonov05e}) as
\begin{equation} \label{theq:6}
\xi^2 = 2\lambdabar_\mathrm{C} r_0 \lambda N = \frac{3}{4\alpha\pi}
\sigma_\mathrm{C}\lambda_\mathrm{las}N_\mathrm{las} \approx
32.7\times N_X \; .
\end{equation}

Here $\alpha $ is the fine-structure constant, $r_0$ the classical
electron radius, $N_X$ number of gammas scattered by each electron
in the bunch when travelling over a wavelength of laser radiation.

Thus, the radiation begins to transform into the synchrotron
spectrum if the electron scatters more than 3 quanta per 100 laser
wavelengths.The linear Compton (Thomson) scattering takes place when
$\xi^2 \ll 1$.

Quantum effects -- recoil of the electron when scattering off the
laser photon which lead to change the cross section and spectrum --
are governed by the parameter
\begin{equation} \label{theq:7}
a=2(1+\cos\varphi)\frac{\gamma E_\mathrm{las}}{m_e c^2} =
\frac{E_X^\mathrm{max}}{E_\mathrm{beam}}\;.
\end{equation}

If $a\ll 1$ then no quantum effects such as decreasing of the cross
section and of the energy of scattered gamma-ray quantum are
appeared.

\subsection{Estimations} Generation of the required amount of
gamma--ray quanta imposes certain demands upon number of laser
photons scattered off the electron per a passage through all 30 IPs
(IPs section): Each of $6.24\times10^{10}$ should in average scatter
off $0.223$ of gamma--ray quanta with the energy within $23.2\ldots
29$\,MeV interval. Since the total length of all IPs is $30\times
0.9 \times 2.5 = 67.7$\,mm\,$\approx 6770$ wavelengths of CO2 laser
$\lambda\approx 10\,\mu$m, then the condition of linear Compton
scattering fulfils. For the case considered -- $E_X=29$\,MeV,
$E_\mathrm{beam}=4.1$\,GeV -- the parameter $a \approx 10^{-2}\ll
1$. Thus, the conditions for the linear Compton (see, e.g., Thomson
\cite{thomson1906}) are fully satisfied.

For the linear Compton spectrum, within the energy interval of $23.2
\ldots 29$\,MeV there emitted are $0.248$ of the total number of
photons.%
\footnote{These photons transmit about 0.45 of the gamma-beam
energy. Hence, only the half of energy is useless and has to be
dumped.}
The required number of photons is bigger than estimated above, about
$0.9$ of gammas per electron's passage through IPs section.
Nevertheless, the conditions for linear Compton scattering are still
satisfied with good margins.

\begin{figure} [hbtp]
\centering
\includegraphics[width=0.8\textwidth]{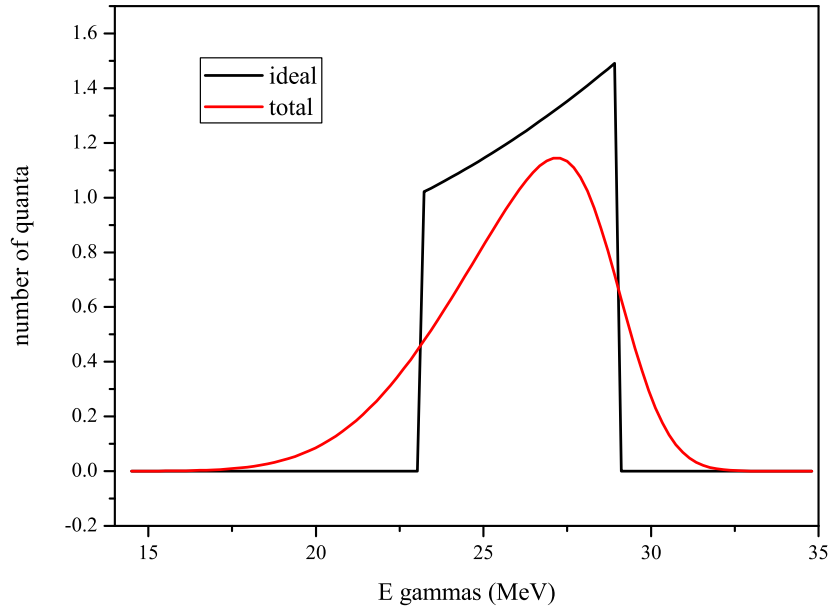}%
\caption[]{Collimated spectra of gammas. Black curve for the ideal
case, red for simulated \label{spectrum29}}
\end{figure}

The spectrum of collimated gamma rays in small-angle approximation
(see \cite{buskoepac04}) is presented in Fig.\ref{spectrum29}. The
collimator was adjusted to cut out gammas with energy below $23.2$
for the ideal case of parallel electron trajectories and zero energy
spread.

\subsection{Compton ring dynamics}
Statistical properties of the bunches circulating in Compton storage
rings are mainly governed by the synchrotron radiation and
scattering of laser photons off the electrons.

Being statistically independent, these two processes establish the
stationary properties of the bunches -- emittances, squared energy
spread $s^{2}$, or squared bunch length $\sigma_y^2$ -- as (see
\cite{buepac04})
\begin{equation} \label{eqxlo1}
s^{2}   = \frac{ s_{s}^{2}\Delta E_{s}+s_{X}^{2}\Delta E_X}{\Delta
E_{s}+\Delta E_{X}}\; ,
\end{equation}
where $s_{X}=\left. \Delta E_\mathrm{beam}\right/E_\mathrm{beam}$ is
the partial Compton energy spread (rms).

The squared partial Compton energy spread and rms bunch length read
\begin{eqnarray}
s_{x}^{2}  &  = &\frac{E_X^\mathrm{max}}{6E_\mathrm{beam}}\; ;\label{eqxlo2}\\
\sigma_y^2 &= &\frac{\alpha_1 E_X^\mathrm{max} h w^2}{12\pi e
V_\mathrm{rf}}\; , \label{eqxlo3}
\end{eqnarray}
with $w \equiv c\left/ f_\mathrm{rf}\right. $ being the width of rf
bucket (for 650\,MHz $w=0.461$\,m), $\alpha_1$ the linear momentum
compaction factor. As it should be emphasised, the partial Compton
energy spread for a given spectrum is determined only by the energy
of circulating electrons.

Energy loss of the electron for the rings under consideration is
$\Delta E_X=0.9\times 29/2 = 11.2$\,MeV per passage (or turn).

Synchrotron losses depend on the radius of curvature $\rho$ in the
bending magnets and the energy of electrons.
The radius of curvature is
\[
\rho=\frac{E_\mathrm{beam}}{0.3B}%
\]
where $\rho$ in meters, $B$ is the dipole field strength in Tesla,
$E_\mathrm{beam}$ the energy of electrons in GeV.
Energy losses $\Delta E$ (in keV) per turn is
\[
\Delta E_s=\frac{88.5E^{4}}{\rho}=26.55 B E_\mathrm{beam}^{3}\, .
\]

Hence, for $E_\mathrm{beam}=4.1$ GeV and $B=0.63377$\,Tesla every
electron looses $1.160$\,MeV/turn which is sufficiently smaller as
compared with the Compton losses ($11.2$\,MeV/turn, see above). The
beam dynamics in CO2 ring will be laser-dominated, the rms
steady--state energy spread (\ref{eqxlo2}) $\left. \Delta
E_\mathrm{beam}\right/E_\mathrm{beam} = 3.43\times 10^{-2}$. Hence,
to provide stable lossless circulation of the beam, a
low--compaction optics is almost mandatory \cite{gladkikh05}.

For the listed in Table~\ref{listpara} parameters and the extremely
low compaction optics (see \cite{feikes04}), $\alpha_1=2\times
10^{-6}$, the partial Compton stationary bunch length is $4.85$\,mm.
Computation of steady--state yield for CO2 ring with
$\alpha_1=2\times 10^{-6}$ and other parameters from
Table~\ref{listpara} by an analytical code based on (\ref{eqxlo1})
for all three degrees of freedom results in following: the yield per
electron passage through IPs is $0.43$, the bunch length $4.53$\,mm,
the horizontal emittance reduced down to $8.64\times
10^{-11}$\,rad\,m. Thus, number of $(23.2\dots 29)$\,MeV gammas is
$6.2346\times 10^{9}$.

Summarising the feature of steady--state regime of Compton ring
operation, it has to be stated that this regime is unacceptable due
to the following reasons:
\begin{enumerate}
\item very high RF voltage is required to provide stable motion: if
average energy losses per turn exceeds RF amplitude, the RF bucket
vanishes, see eq.~(\ref{eq:stackone}) for $q\le 1$;

\item even if RF voltage is high enough, fluctuations in number and
energy of emitted gammas results in very short beam life time (i.e.
high quantum losses);

\item and even the short life time could be regarded as tolerable,
bunch lengthening up to the steady--state 
will reduce the
gamma yield below demands.

\end{enumerate}

\subsection{Pulsed mode: basic idea}
Consider a ring with very low synchrotron number $Q_s$ such that the
inverse to it -- the period of synchrotron oscillations -- much
exceeds the temporal length of gamma burst $T_X$: $T_X\ll 1/Q_s$.
Under such condition we can consider the bunch as free, moving in
the phase plane straight with constant velocity. (Transformation of
the energy spread into the bunch length takes at least a quarter of
the synchrotron period.) Interactions of the bunch with the laser
pulse (recoils) result in Brownian dispersion and offset along the
momentum axis $p$ (here and below $p$ is relative deviation of the
energy of electron $E_e$ from the synchronous $p\equiv
(E_e-E_\mathrm{beam})/E_\mathrm{beam}$): The dispersion and offset
after $n$--th turn are $\sigma_p^2(n)=n\left<\Delta p^2\right>$ and
$S_\mathrm{dr} = n\left< \Delta p\right>$, respectively. The offset
is caused by drift due to the fact that any interaction of the
electron with a laser pulse results decreasing $p$.

Thus, if we set a bunch at the location in the phase space where the
phase velocity directs vertically (along $p$--axis) and is equal to
the drift velocity opposite to it, then the bunch during the burst
will not change location of its centre-of-weight and length.

The scheme will be improved if the curvature of phase trajectories
at the initial bunch location is low, which can be attained for the
nonlinear ring optics with sufficient cubic term in the phase slip
factor.

\subsection{Simulations}

Simulations were performed with a code intended for modelling the
longitudinal dynamics in Compton rings with nonlinear compaction
\cite{buglasko05}. The code represents a turn constructed from 4
sections:
\begin{enumerate}
\item IPs -- a particle can experience a series of random kicks,
probability of which conforms with the particle phase coordinate
$\phi$ and 3-D Gaussian distributed photon population in laser
pulse; the random amplitude of the kick obeys the linear Compton
spectrum.

\item
SL -- losses due to synchrotron radiation: every particle
experiences the regular kick with amplitude proportional to the
particle instant energy $p$.

\item
DRIFT -- the phase of the particle advances as $\phi_f = \phi_i +
\kappa_1p + \kappa_2p^2 + \kappa_3p^3$ ($\kappa_{j}$ are the
normalised phase slip factors).

\item
RF -- zero--length harmonic rf cavity: every particle receives a
kick with amplitude depending of its phase, $p_f = p_i -
U_\mathrm{rf}\sin\phi $.

\item
STAT -- in--flight statistics of the bunch.

\end{enumerate}

The particle in the bunch is represented by the flat zero--length
disk with 2-Dim density distribution.

First simulations were performed to validate basic idea of the
pulsed mode. The simulations have proved our expectations true.

Then simulations of the both CO2 and YAG rings were performed. For
each of the rings two optics were chosen: a conventional linear
(`lin') and a low--alpha nonlinear (`nl') with compaction factors
like \cite{feikes04}.
The quadratic compaction factors, $\alpha_2$, were chosen to provide
zero value to the quadratic phase slip factors $\kappa_2$, see
\cite{yang99}.
For every optics, the initial position of the bunch in the phase
plane were being set manually to attain about a maximal average
yield over the given number of turns.

Results are summarised in Table~\ref{listres}.

\begin{table}[hbtp]
\caption{Results of simulations\label{listres}}
\centering 
\vspace{2pt}
\begin{tabular}{|l|r|r|r|r|}
\hline parameter  & CO2 con & CO2 nl& YAG con & YAG nl \\[1ex]
\hline Linear comp $\alpha_1$ &  $1\times 10^{-4}$ & $ 2\times
10^{-6}$ & $1\times 10^{-4}$ & $2\times 10^{-6}$ \\[1ex]
NL comp $\alpha_2$  & $-2\times 10^{-4}$ & $-4\times 10^{-6}$ &
$-2\times 10^{-4}$ & $-4\times 10^{-6}$  \\[1ex]
NL comp $\alpha_3$  & $0$&$2\times 10^{-3}$ & $0$&$2\times 10^{-3}$\\[1ex]
RF voltage ( MV )  &20 & 20 &20 & 20 \\[1ex]
Harmonic number  & $1408$ & $1408$ & $600$ & $600$\\[1ex]
Turns laser on  & $50$ & $50$ & $100$ & $100$\\[1ex]
Aver. gammas/electron &  $0.39$ & $1.24$&  $0.21$ & $0.94$\\[1ex]
[23.2-29]\,MeV gam/pulse &  $5.65\times 10^{9}$ &
$1.78\times 10^{10} $ &  $3.0\times 10^{9}$ & $1.36\times 10^{10}$\\[1ex]
\hline
\end{tabular}
\end{table}

There were not lost particles for the both rings in $4500$-turns
operating cycle with the first 50 or 100\,turns with the laser on.

It should be pointed out that the maximal theoretical yields for the
given geometry -- the crossing angle, transversal dimensions of the
bunch and all dimensions of the laser pulse are $2.78\times 10^{10}$
(CO2) and $3.04\times 10^{10}$ (YAG).

\begin{figure} [hbtp]
\centering
\includegraphics[width=0.45\textwidth]{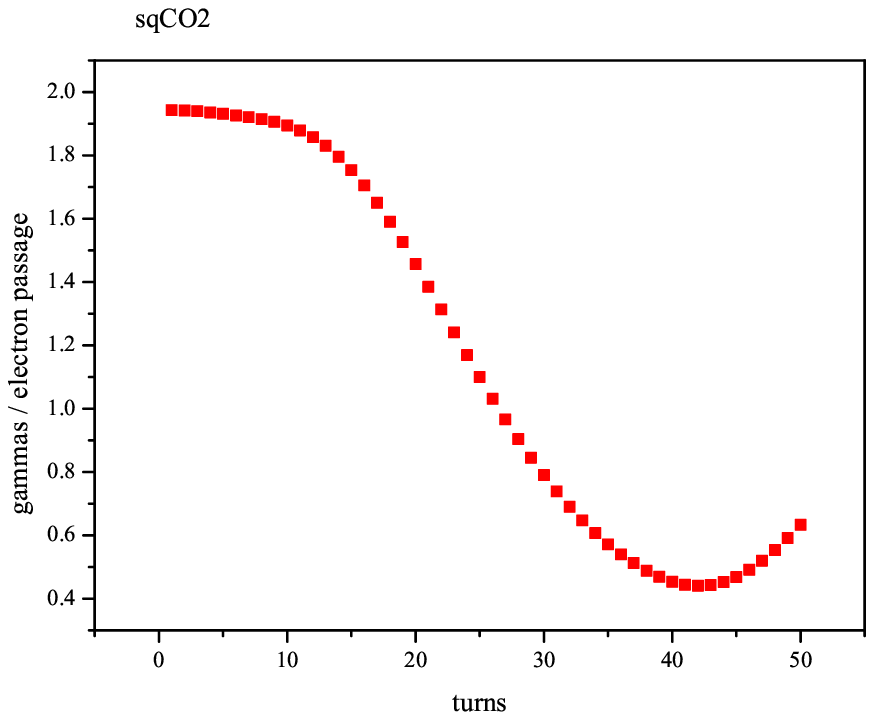} 
\includegraphics[width=0.45\textwidth]{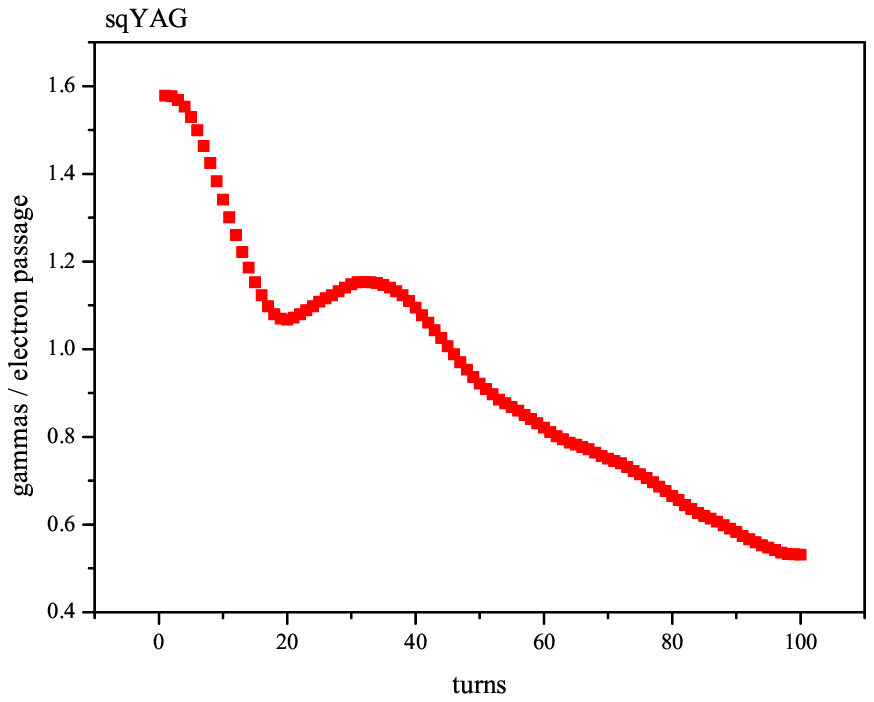}%
\caption[]{Yield vs turn, nonlinear rings. Left -- CO2, right -- YAG
\label{yieldid}}
\end{figure}

Number of scattered quanta per electron per passage through IP
obtained from simulations is presented in Fig.\ref{yieldid}.

\begin{figure} [hbtp]
\centering
\includegraphics[width=0.8\textwidth]{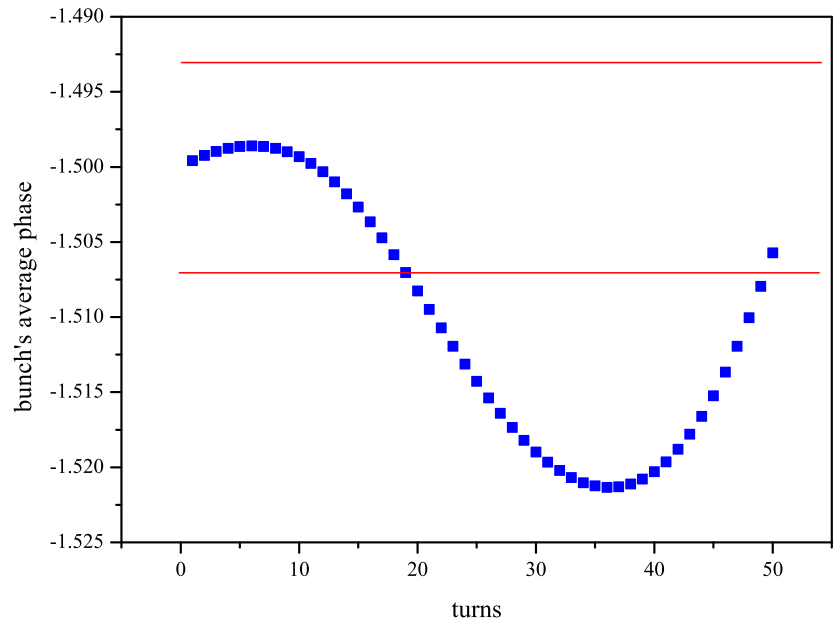}%
\caption[]{Centre-of-weight excursion. Band with red border lines
represent effective rms interaction length \label{centfi}}
\end{figure}

Nonlinear dependence of the yield on time occurs due to deviation of
the bunch centre of weight, as presented in Fig.\ref{centfi}.

As it can be seen, the minimum in yield caused by the bunch offset
from the laser focus.

\begin{figure} [hbtp]
\centering
\includegraphics[width=0.8\textwidth]{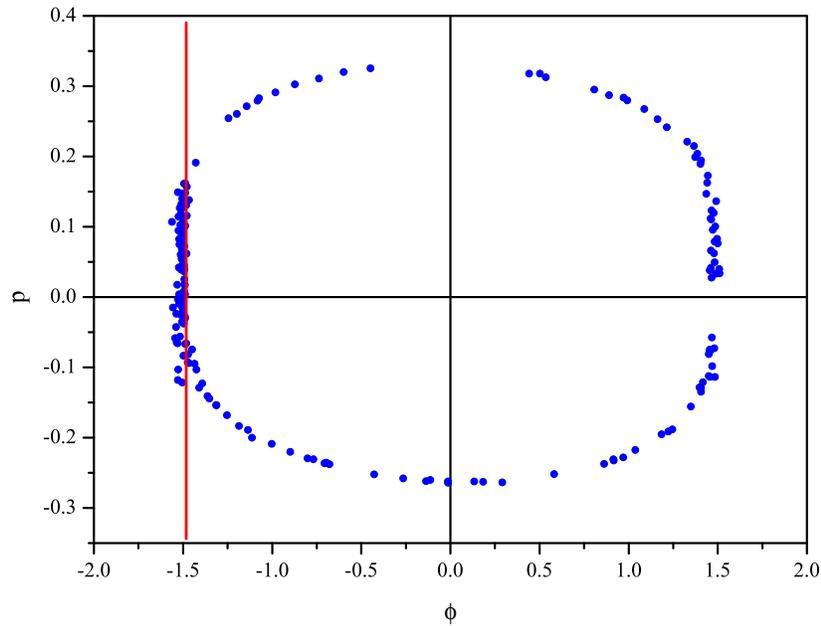}%
\caption[]{Phase distribution of electrons in the bunch after 100
turns \label{phase100}}
\end{figure}

In Fig.\ref{phase100} there presented is distribution of electrons
over the phase space in YAG NL ring after 100 turns. The vertical
red line indicates the phase position of the laser pulse.

\subsection{Conclusion}
The considered systems in the pulsed mode are capable to meet
demands imposed upon the Compton ring with good margins for the
stored laser power and the bunch density. It seems that fine
adjustment of the parameters, especially the ring nonlinearity,
initial bunch position in the phase plane, and laser-to-bunch
matching, can enhance the system performance.


\clearpage
\sect{Beam Stacking in Damping Ring}

The ideal choice of accumulation ring turns out
to be the damping ring itself.
Due to its large circumference, it can store the full
number of positron bunches, while at the same time providing
a significant damping over the 10 ms repetition period of 
the laser-beam collisions in the Compton ring.
In addition, the longitudinal bucket areas of the proposed
damping rings is large (due to the small momentum
compaction factor and high rf voltage), which facilitates
stacking. Choosing the damping ring for accumulation
also avoids constructing another ring.
The damping ring will accumulate for 100 ms and
then damp for another 100 ms (close to 10 damping times),
for a main-linac repetition rate of 5 Hz. 

The damping ring should store 2800 bunches with a bunch spacing
of about 3.077 ns (325 MHz). The corresponding ring 
circumference is about 3.3 km. The beam energy in the damping
ring is chosen as 5 GeV. The beam in the ring 
consists of 10 trains, each with 280 bunches. 
The longitudinal damping time is about 10 ms.

Each injected positron bunch coming from the 
laser-Compton ring has a 
bunch population of $2.4x10^8$. Injection occurs 
into all 2800 buckets on 10 consecutive turns.
This 10-turn injection is followed by a 
20-ms of damping, after which the next 
10-turn injection starts. The whole process
of injection and damping repeats 10 times.
After 10 injections, the bunch charge
is about $2.4\times 10^{10}$ positrons. 

The transverse rms emittance of the injected positrons 
is of order 0.0015 rad-m for an undulator source and 
0.006-0.007 rad-m for a conventional source \cite{klausfloettmann}. 
We expect similar values for a source based on 
Compton back scattering. The ``edge emittance'' and, 
therefore, the required acceptance is about 10 times 
larger than the rms emittance \cite{klausfloettmann}. 
The rms energy 
spread of the injected beam at 5 GeV
is taken to be of order 0.14\% and 
the rms bunch length 3 mm. 
The injected energy spread is close to the
equilibrium value, the bunch length
is about two time shorter.

The rf frequency of the damping ring is taken 
as 650 MHz. then the length of the rf bucket 
is about 45 cm, about 15 times the injected bunch 
length of 3 mm. Therefore, it should be possible to
inject 10 bunches into the bucket without
damping. For a more precise estimate, 
it is instructive to compute the bucket area.  

The maximum energy deviation at the centre of the rf bucket is 
\begin{equation}
\left( \frac{\Delta p}{p_{0}} \right)_{\rm max}^{2}
= \frac{e V_{\rm rf} \sin \psi_{s}}{
\pi h |\eta_{c}| c p_{0}} 2 \left( \sqrt{q^{2}-1}-{\rm
  arccos}\frac{1}{q}\right)\; , 
\label{eq:stackone}
\end{equation}
where $q = eV_{\rm rf}/U_{0}$ is the peak energy gain
from the rf system divided by the energy loss per turn,
and 
\begin{equation}
\psi_{s} = {\rm  arccos} \left(\frac{U_{0}}{e V_{\rm rf}} \right)\; 
\end{equation}
the synchronous phase angle.

The full length of the bucket is given by
\begin{equation}
l_{\rm bucket}  = \frac{\lambda_{rf}}{2 \pi} (\psi_{1}-\psi_{2})\; , 
\end{equation}
with $\psi_{1}$ and $\psi_{2}$ denoting two solutions
of the transcendent equation
\begin{equation}
\cos \psi_{1,2} + \psi_{1,2} \sin \psi_{s} = 
(\pi -\psi_{s}) \sin \psi_{s} - \cos \psi_{s}\; .
\end{equation}
One of the two solutions is $\psi_{1}=\pi -\psi_{s}$.
The second solution $\psi_{2}$ has to be determined
numerically. 

The total bucket area is estimated as 
\begin{equation}
A_{\rm bucket} \approx \pi \; ( \Delta E_{\rm max} )\; 
(\Delta t ) _{\rm max}
\end{equation}
where $\Delta E_{\rm max}\approx c \; \Delta p_{\rm max}$
is the maximum energy deviation at the centre of the bucket,
and $(\Delta t)_{\rm max} = l_{\rm bucket}/(2 c)$,
the bucket half length in units of seconds.

\begin{table}[htbp]
\caption{Example parameters of damping ring
employed for positron stacking.} 
\label{tablestack}
\begin{center}
\begin{tabular}{|lc|}
\hline
energy & 5 GeV \\
circumference & 3323 m \\
particles per extracted bunch & $2.4\times 10^{10}$ \\
rf frequency & 650 MHz \\
number of trains & 10 \\
number of bunches per train & 280 \\
gap between trains (no.~of missing bunches) & 80 \\
particles per injected bunch & $2.4\times 10^{8}$ \\
injections per bucket on successive turns & 10\\
injection repetition rate during 100 ms 
& 100 Hz \\
total number of injections & 100 \\
store time after 100 injections & 100 ms \\
energy loss per turn & 5.5 MeV \\
damping time & 10 ms \\  
transverse emittance at injection & 0.005 rad-m \\
rms bunch length at injection & 3 mm \\
rms energy spread at injection & 0.14\% \\
final rms bunch length & 6 mm \\
final rms energy spread  & 0.14\% \\
longitudinal ``edge'' emittance at injection & ~0.7 meV-s \\
rf voltage & 20 MV \\
momentum compaction & $3\times 10^{-4}$\\
2nd order momentum compaction & $1.3\times 10^{-3}$\\
synchrotron tune & 0.0365 \\ 
bucket area & 292 meV-s \\
synchronous phase & 15.58$^{\circ}$ \\
separatrix phases 1 \& 2 & $164.42^{\circ}$, $-159.19^{\circ}$\\ 
maximum momentum acceptance & $\pm 2.7$\% \\
\hline 
\end{tabular}
\end{center}
\end{table}

We have simulated the efficiency of the
stacking process. The simulation considers only
the longitudinal phase space. It includes
a sinusoidal rf voltage, and the first
and second order momentum compaction
factors. Radiation damping and quantum
excitation are modelled as in \cite{siemann}.
Damping-ring and beam parameters 
assumed for this simulation are listed in 
Table~\ref{tablestack}. They are similar
to those of the 3-km ILC
damping ring designs PPA and OTW
\cite{wolski}. 

The injection septum is
placed at a location with large
dispersion, since the stacking
is performed in longitudinal
phase space. The septum blade
is assumed to be small compared
with the transverse 
dispersive beam size. 
Between successive turns
of injection, the orbit at
the septum is varied with
fast bumper magnets. 
It may prove necessary to install injection
septa on either side of the beam pipe,
and to move the orbit from one side to
the other after 5 from 10 injections,
so as to facilitate the injection of
bunches with opposite sign of
relative momentum deviation and
optimum phase-space coverage over
10 turns. 

The energy of the
injected beam is ramped such that
the transverse position 
of the septum always corresponds to
a separation of $2 \sigma_{\delta}$ 
from the beam centroid,
taking into account the local
dispersion value. 
Consequently, 
positrons of the stored beam  
being closer than $2\sigma_{\delta}$
to the injected beam at the turn of
injection are considered lost.
Likewise, injected positrons with a momentum 
offset of more than $2\sigma_{\delta}$
from the injected beam centroid energy,
in the direction towards the bucket centre,
are also taken to be lost. 
The initial energy offset is first
changed from $-1.05$\% to $-2.45\%$,
in steps of $-0.35\%$ from turn to turn, 
and then from $+1.05$\% to $+2.45\%$
in steps of $+0.35\%$,
over a total number of 10 turns.
The point $\delta=0$ is left out, since this
equals the position of the accumulated
damped beam.

\begin{figure}[htbp]
\begin{center}
\rotatebox{-90}{\scalebox{0.18}{\includegraphics*{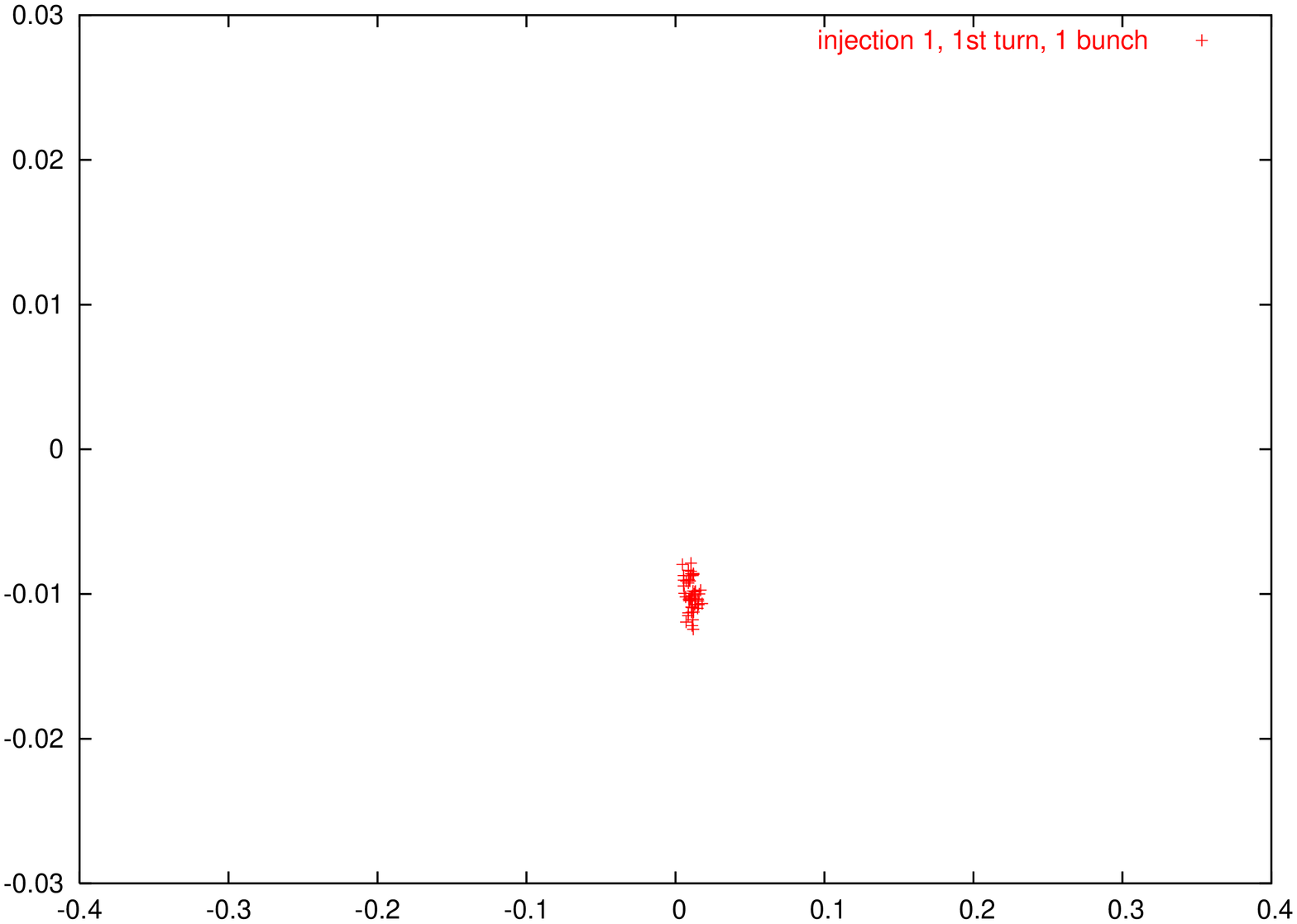}}}
\rotatebox{-90}{\scalebox{0.18}{\includegraphics*{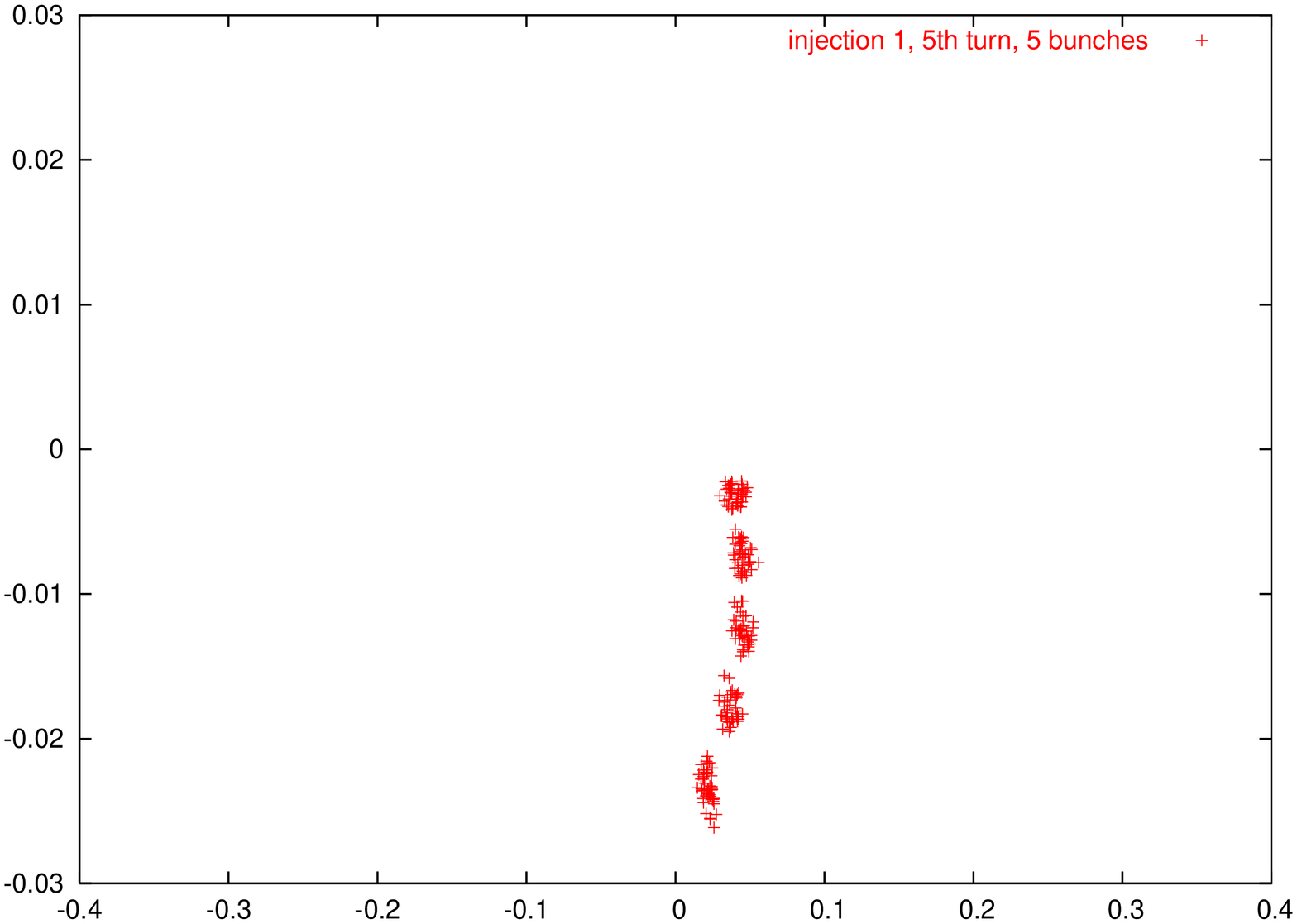}}}
\rotatebox{-90}{\scalebox{0.18}{\includegraphics*{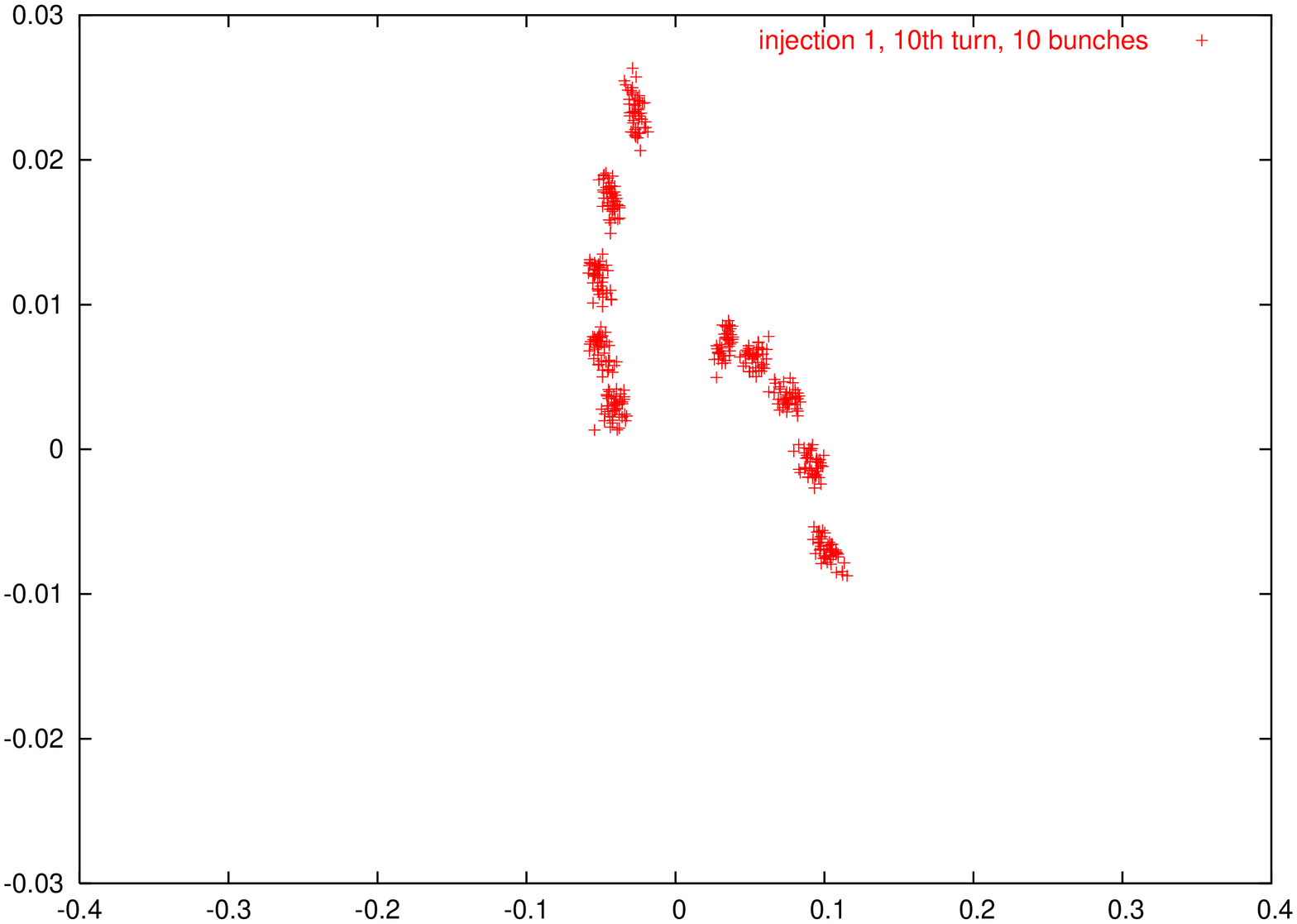}}}
\rotatebox{-90}{\scalebox{0.18}{\includegraphics*{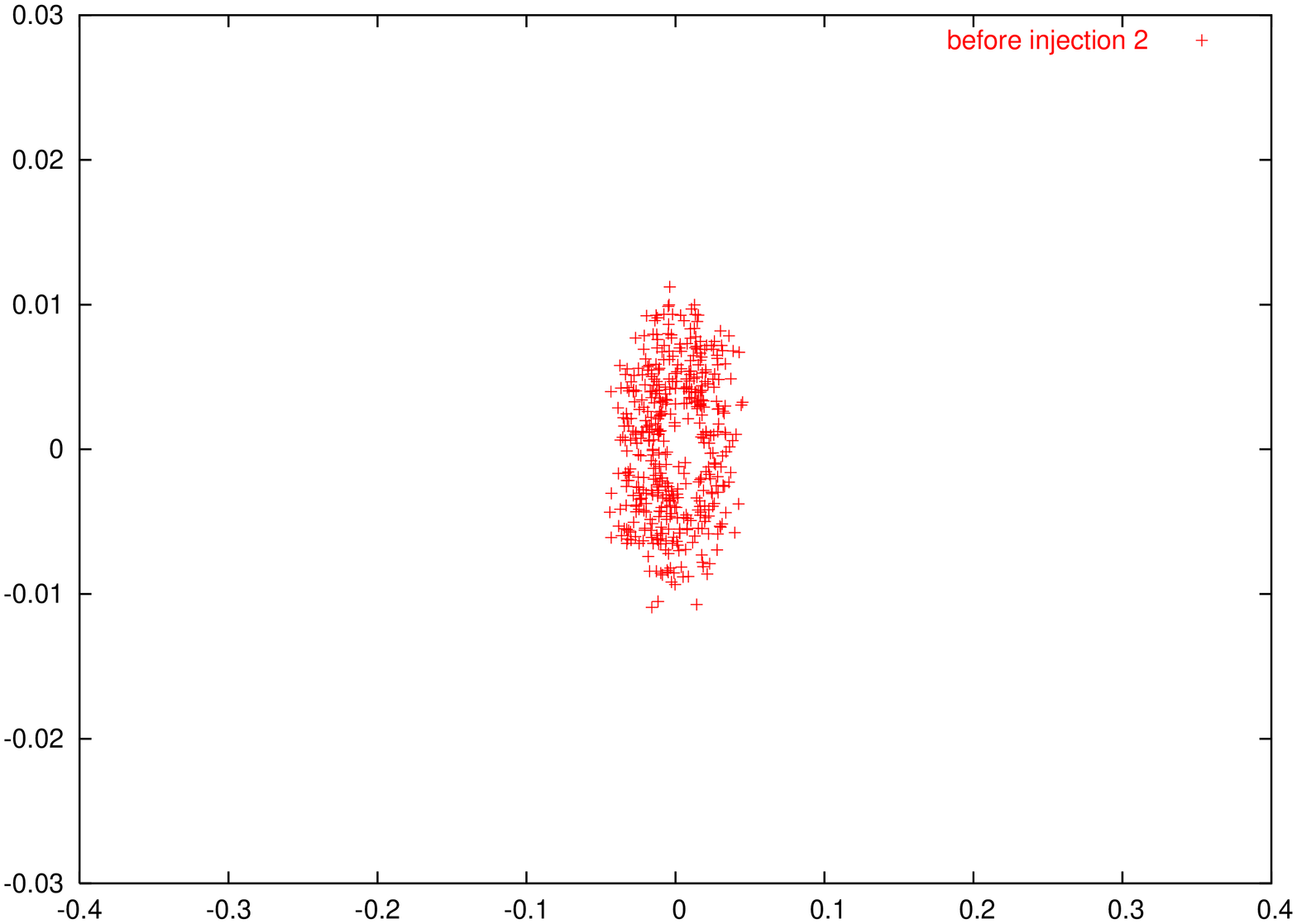}}}
\rotatebox{-90}{\scalebox{0.18}{\includegraphics*{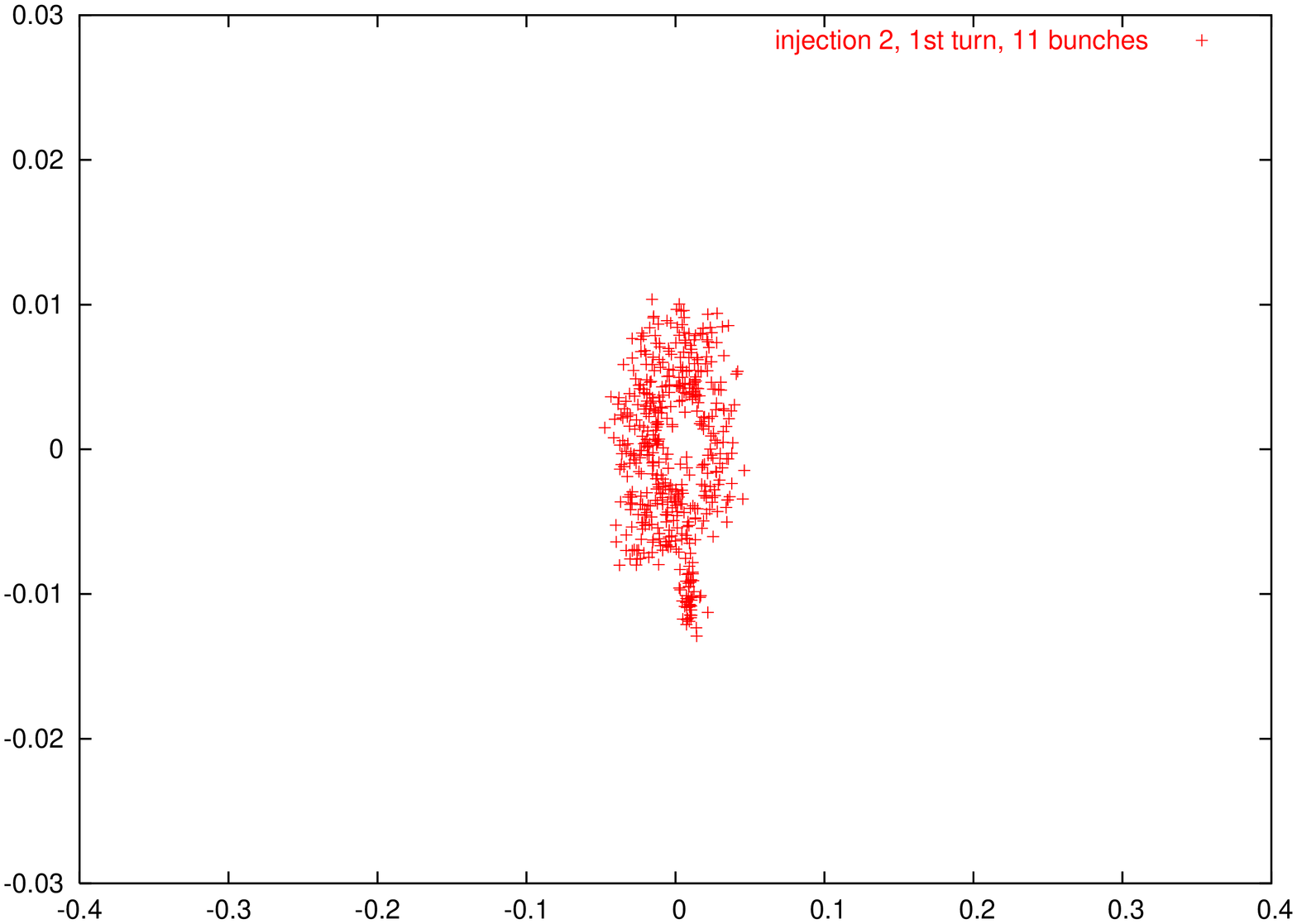}}}
\rotatebox{-90}{\scalebox{0.18}{\includegraphics*{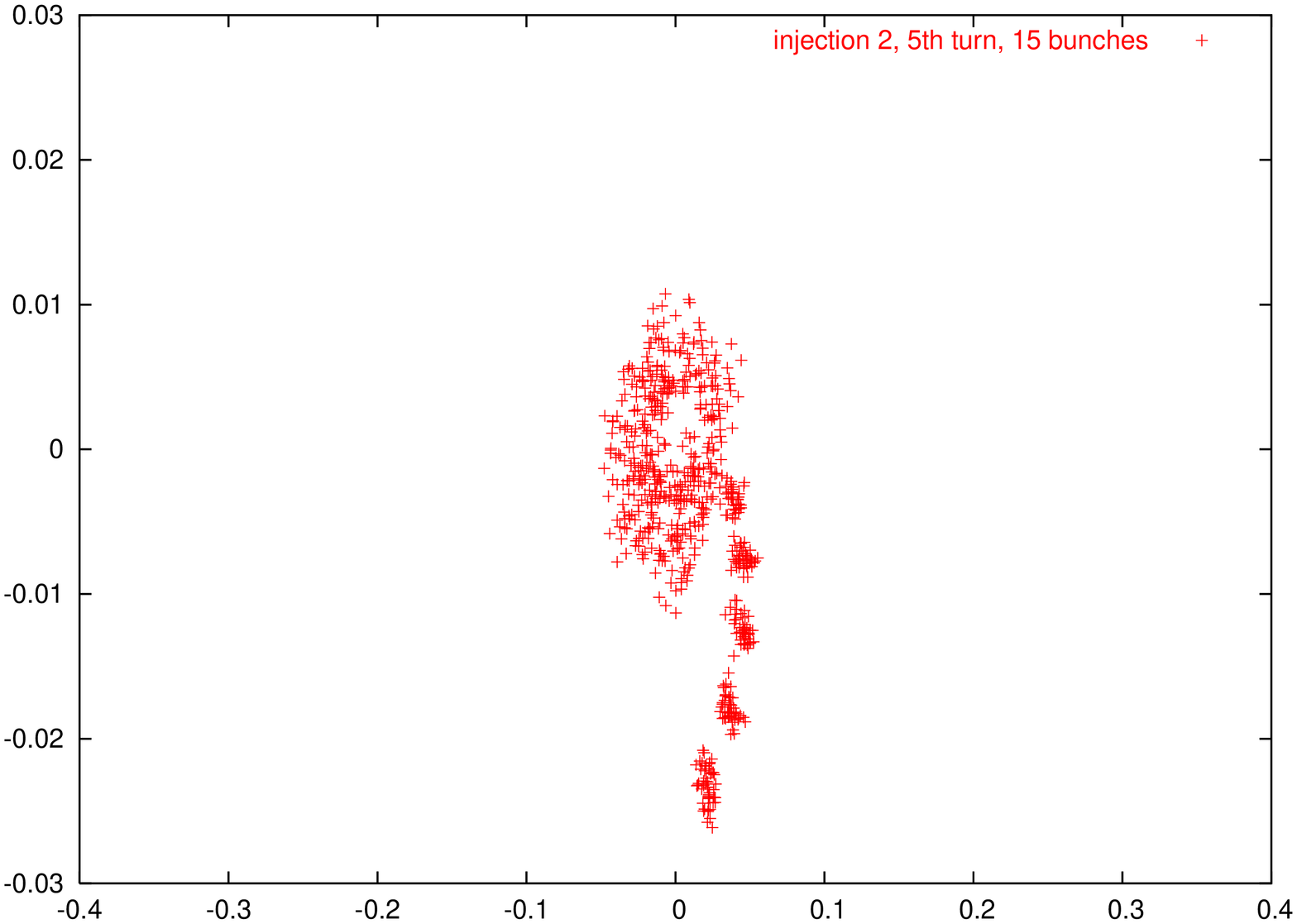}}}
\rotatebox{-90}{\scalebox{0.18}{\includegraphics*{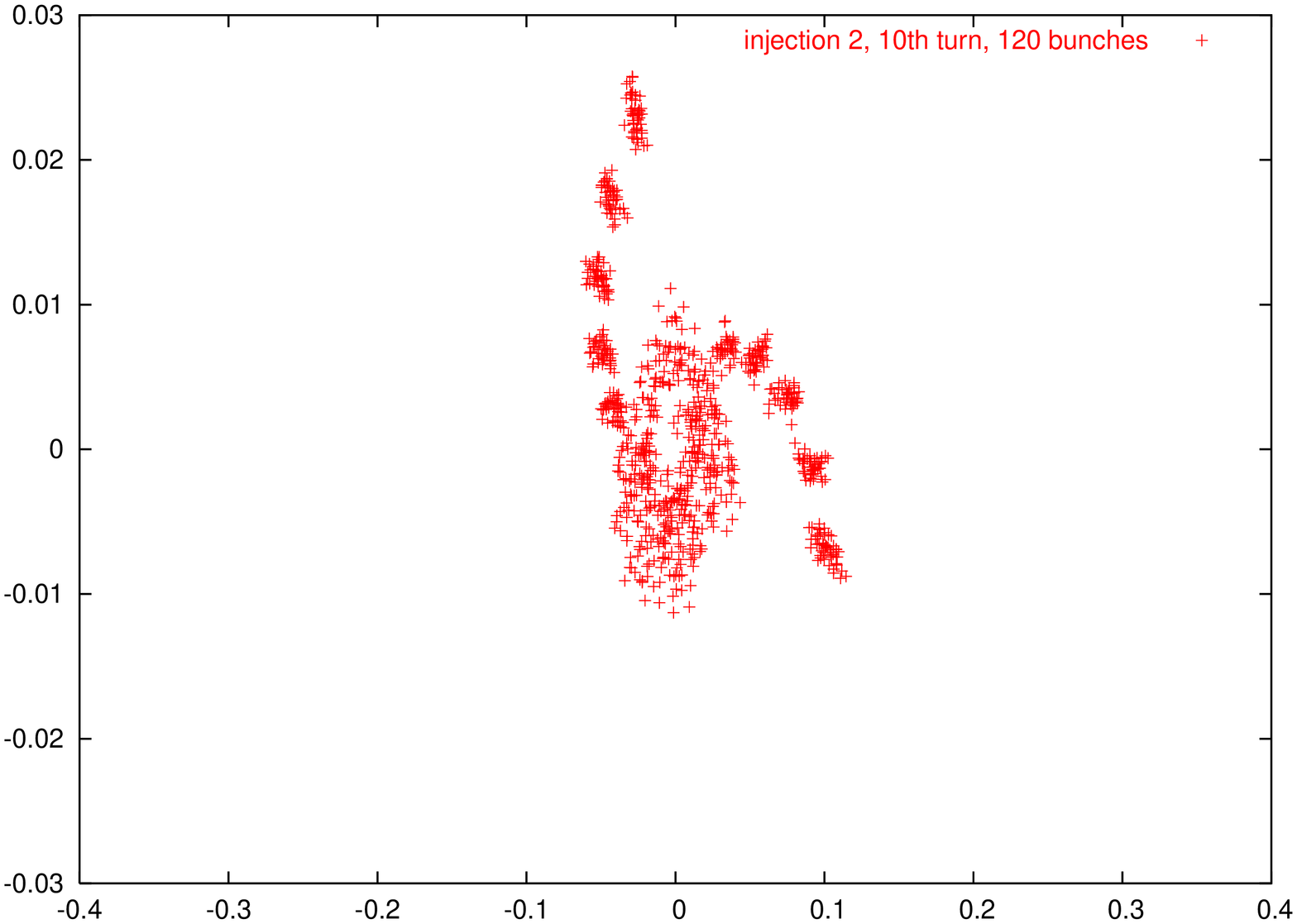}}}
\rotatebox{-90}{\scalebox{0.18}{\includegraphics*{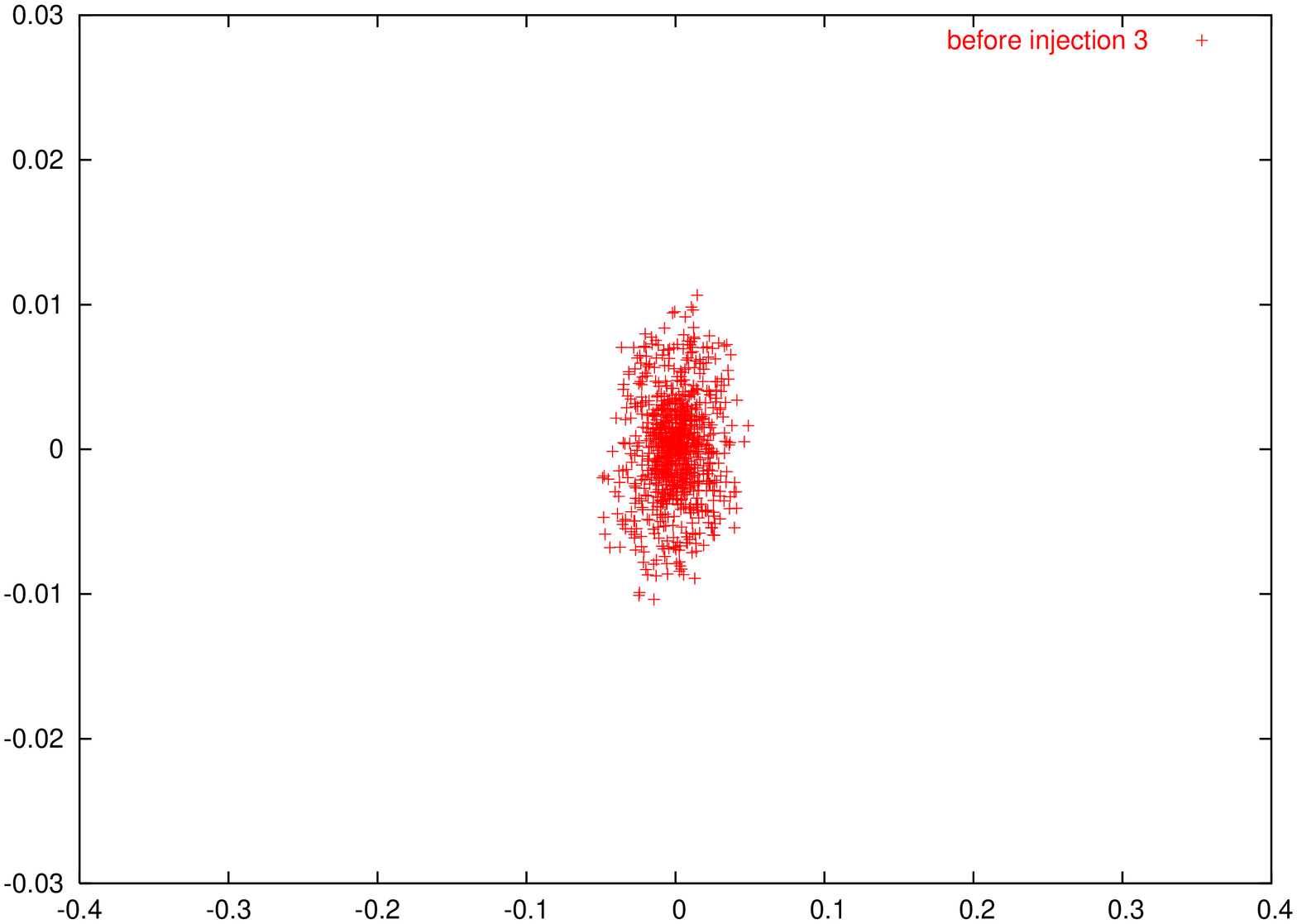}}}
\rotatebox{-90}{\scalebox{0.18}{\includegraphics*{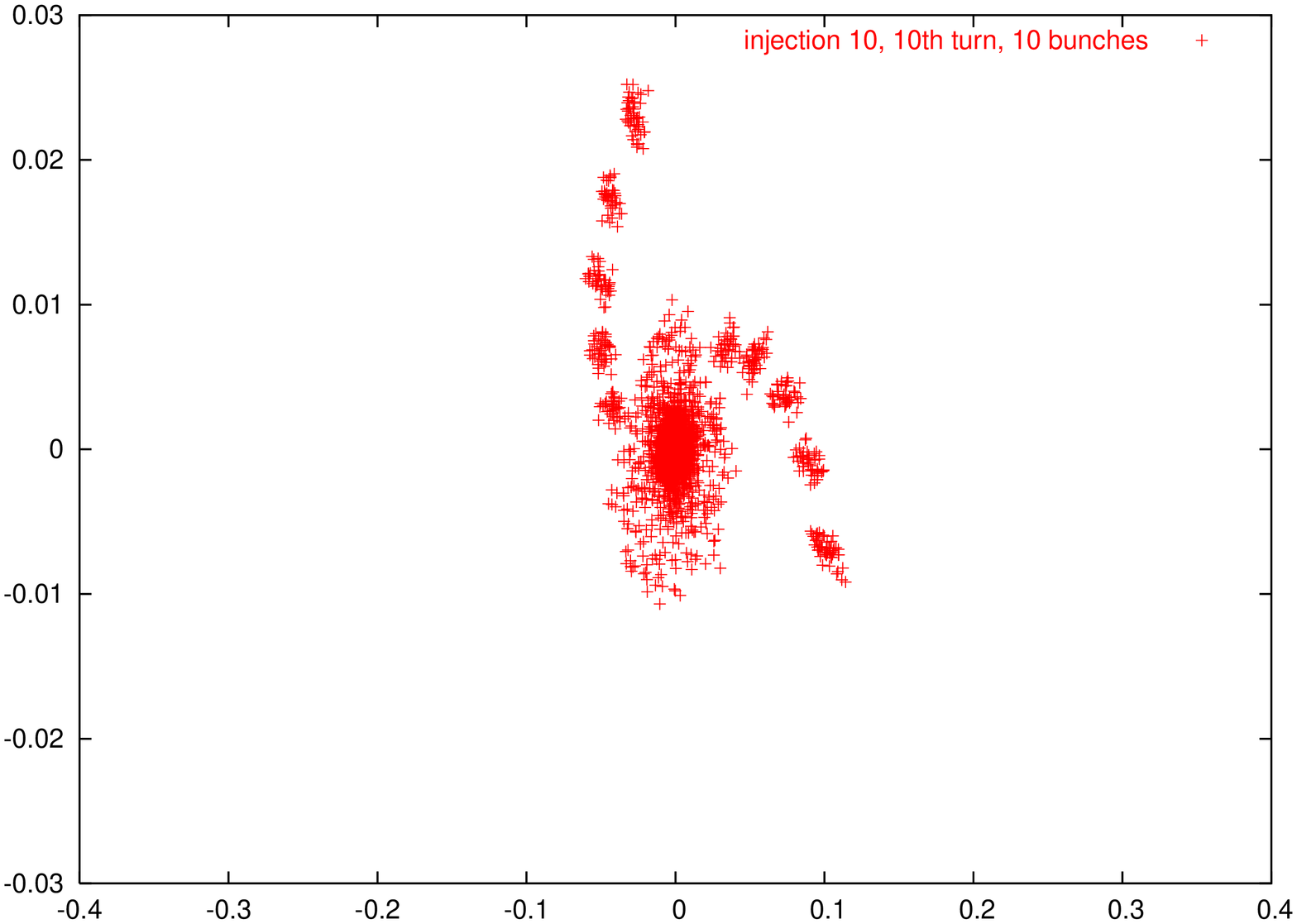}}}
\rotatebox{-90}{\scalebox{0.18}{\includegraphics*{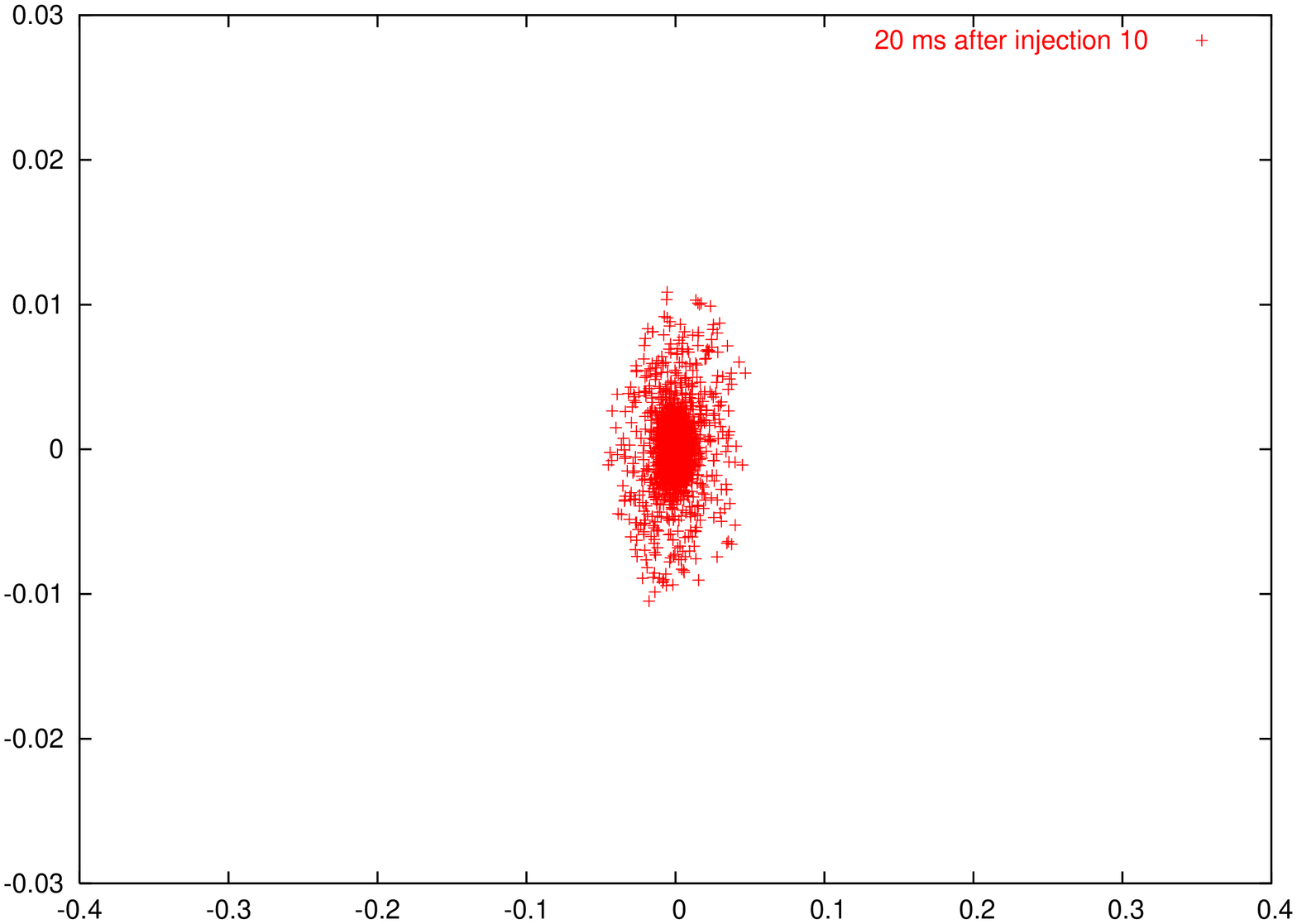}}}
\rotatebox{-90}{\scalebox{0.18}{\includegraphics*{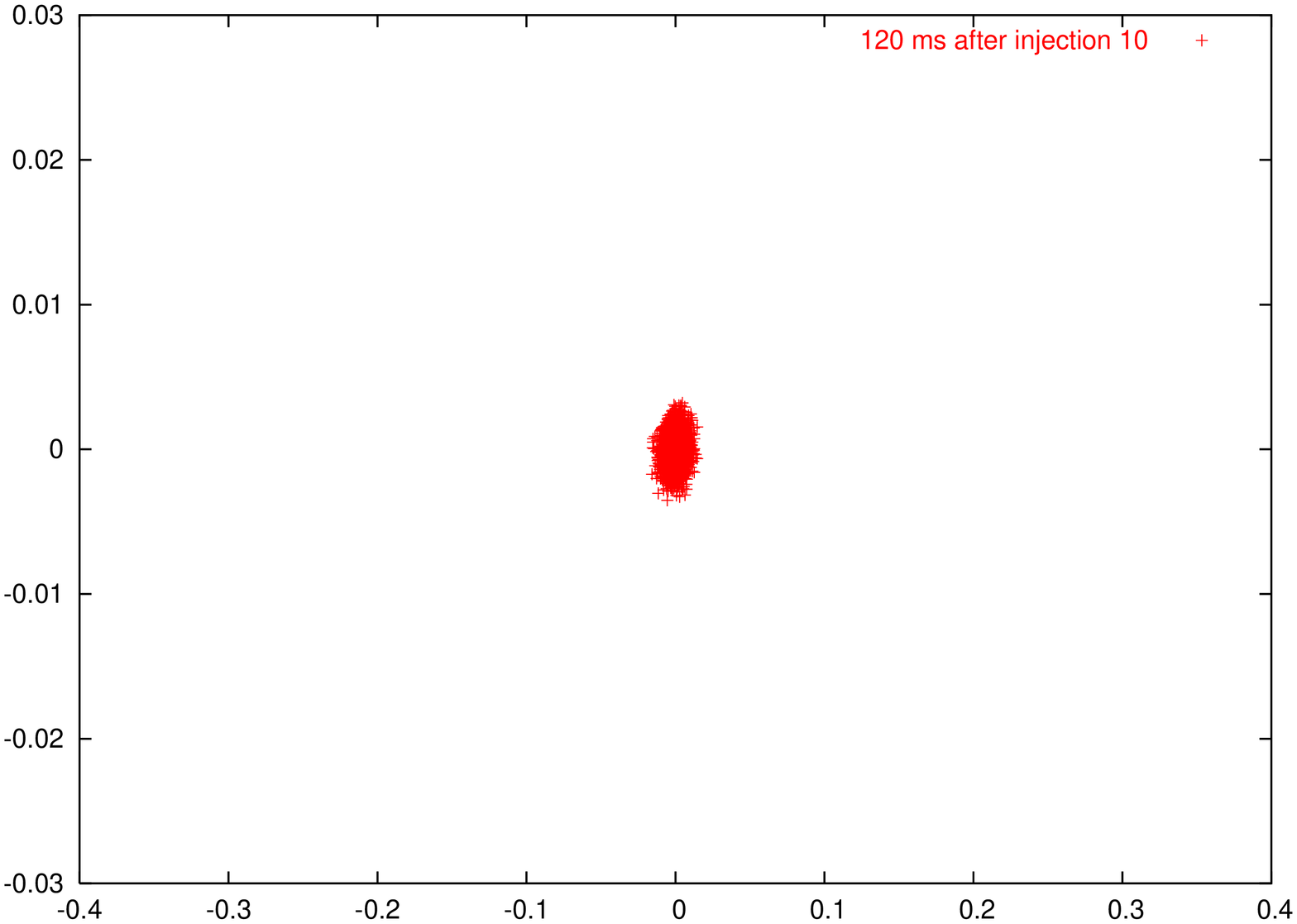}}}
\end{center}
\caption{Snapshots of longitudinal phase space
during injection and stacking:
(a) 1st bunch on 1st turn,
(b) 5th bunch on 5th turn,
(c) 10th bunch on 10th turn,
(d) before 11th bunch on 941st turn,
(e) 11th bunch on 942nd turn,
(f) 15th bunch on 946th turn,
(g) 20th bunch on 951st turn,
(h) before 21st bunch on 1882nd turn,
(i) 100th bunch on 8479th turn,
(j) 100 bunches on 9410th turn,
(k) 100 bunches on 18820th turn.
}
\label{figstack}
\end{figure}

With these assumptions, the  simulated
total loss during accumulation is
about 18\%. Figure \ref{figstack} shows 
snap shots of the accumulation 
process in the damping ring.

\clearpage
\sect{Laser System Design}





\begin{figure*}[hbtp]
\includegraphics*[width=\linewidth]{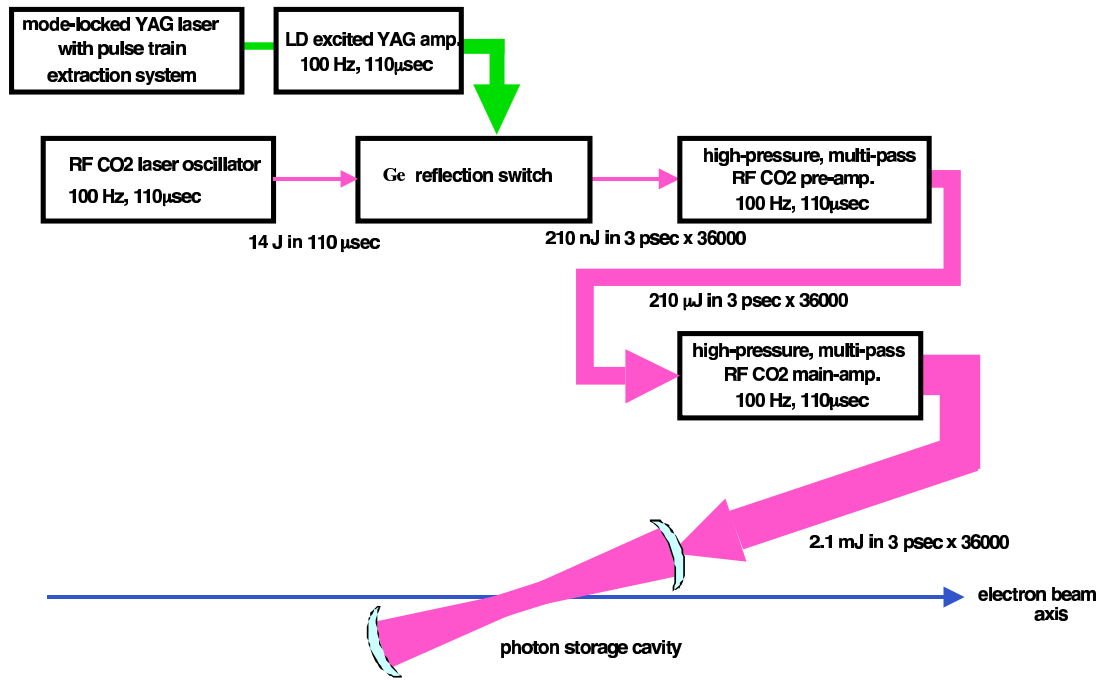}
\caption{\label{CO2-laser}Schematic design of a CO2 laser system.}
\end{figure*}

Figure~\ref{CO2-laser} shows a schematic design of a high-power multi-bunch
CO$_2$ laser system.  A long, $\sim110 \mu $s low-energy pulse is  
generated by a CO$_2$ laser oscillator operating at 100 Hz.
The energy of this long pulse is 14J.
To realize such a long pulse, the CO$_2$ oscillator is
pumped by an RF pulse.

This long pulse is sliced into $\sim3.6\times 10^4$ short, 3ps in rms, bunches
(seed bunches) by Germanium ({\bf {Ge}} )-plates \cite{Slice}, which
can switch between reflection and transmission.
Each short bunch contains energy of 210nJ.

The reflectivity/transparency of the {\bf {Ge}} -plates is
controlled by a multi-bunch pico-second YAG laser.
The YAG laser consists of a mode-locked laser oscillator and
an LD pumped amplifier.

After slicing, $\sim3.6\times 10^4$ CO$_2$ seed bunches are amplified by a
preamplifier and a main amplifier. 
Gains of the preamplifier and the main amplifier are 1000 and 10, respectively. 
Both amplifiers are RF-pumped in order to realize long pulse ($\sim110 \mu $s).
The preamplifier operates at 100 Hz.
For the main amplifier, interleaved 100 Hz operation will be
adopted, where the amplifier operates at 100 Hz during 100 ms, and then takes
a rest for the next 100 ms.  This interleaved mode of operation reduces
the average beam power and, therefore, the heat load on the main amplifier.

Finally, we obtain $\sim3.6\times 10^4$ bunches, each of which contains 2.1 mJ
and has a bunch duration of $\sim 3$ ps in rms.
This short bunch width is necessary in order to match the laser-pulse-length
to a depth-of-focus of $\sim 1$ mm for the parabolic-mirrors,
which create the final small spot size.
To achieve such a short bunch width, the pressure of the CO$_2$ gas in
the pre and main amplifiers must be high (10 atm.).

In the case of a YAG laser system, we need a long, $\sim92 \mu $s, 
low-energy pulse generated by a 325-MHz modelock 12-W YAG laser 
oscillator with 10-ps (FWHM) pulse width. 
To amplify the laser pulse energy from 37 nJ to 6 mJ, four 
pass amplification is necessary with 
a long-pulse flash lamp ($\sim100 \mu $s) and with an 
appropriate cooling system for the flash lamp and the 
YAG rod. 

At this stage, we do not know whether a 100-kW YAG laser system 
is available or not, but the present laser system of the ATF 
photo-cathode RF gun can generate about $0.1mJ/pulse$ over 400
pulses from a 357-MHz 6-W laser oscillator by a two-pass 
amplification. We are planning to increase the pulse train length 
from $1\mu s$ to $50\mu s$ and implement a 4-pass 
amplification system. 
An ultimate technical limit on the maximum power is 
set by thermal effects on the various elements. 
If 100-kW YAG operation turns out to be impossible, 
we have to increase the enhancement factor of the optical
cavity from 100 to about 1000. In any case required
are a detailed engineering design of the laser system
and an experiment on the generation of a huge laser power 
in the burst operation mode.

If otherwise we want to refer to the existing thin-disk laser head technology
(saturable absorber) \cite{thindisk} and associate it with high finesse
cavities ( gain 10000 - see appendix \ref{sec:rdorsay}) we will anyway require
an enhancement of a factor $\sim 50$ in flux but this scheme has the advantage
to work in continuous mode and so with full duty cycle. Therefore the
remaining enhancement could be attained working on different parameters:
\begin{enumerate}
\item Laser power. The thin-disk laser head technology is very promising and
  could attain even more power for a single laser in the future.
\item Gain of the cavity. The R\&D illustrated in appendix \ref{sec:rdorsay}
  is foreseen to attain a gain of 10000 in the first phase but a consecutive
  enhancement of an order of magnitude can be foreseen.
\item The  increase of  the charge/bunches.
\item Increasing the duty cycle of the positron injection in the damping ring.
\end{enumerate}




\clearpage
\sect{Compton Collision Chamber Design}

We assume the basic beam parameters for the ILC 
main damping ring listed in Table \ref{tablestack}. 
Although there still many parameter options 
for the main damping ring are being discussed, 
our scheme of polarised positron generation 
can be adapted to each option by an appropriate 
change of the laser system, the Compton-ring design 
and the 5-GeV injector linac. 
For simplicity, in this section we describe 
the design of the Compton collision chamber 
only for the YAG laser case.

\begin{figure}[hbtp]
\centering
\includegraphics*[width=0.7\linewidth,clip]{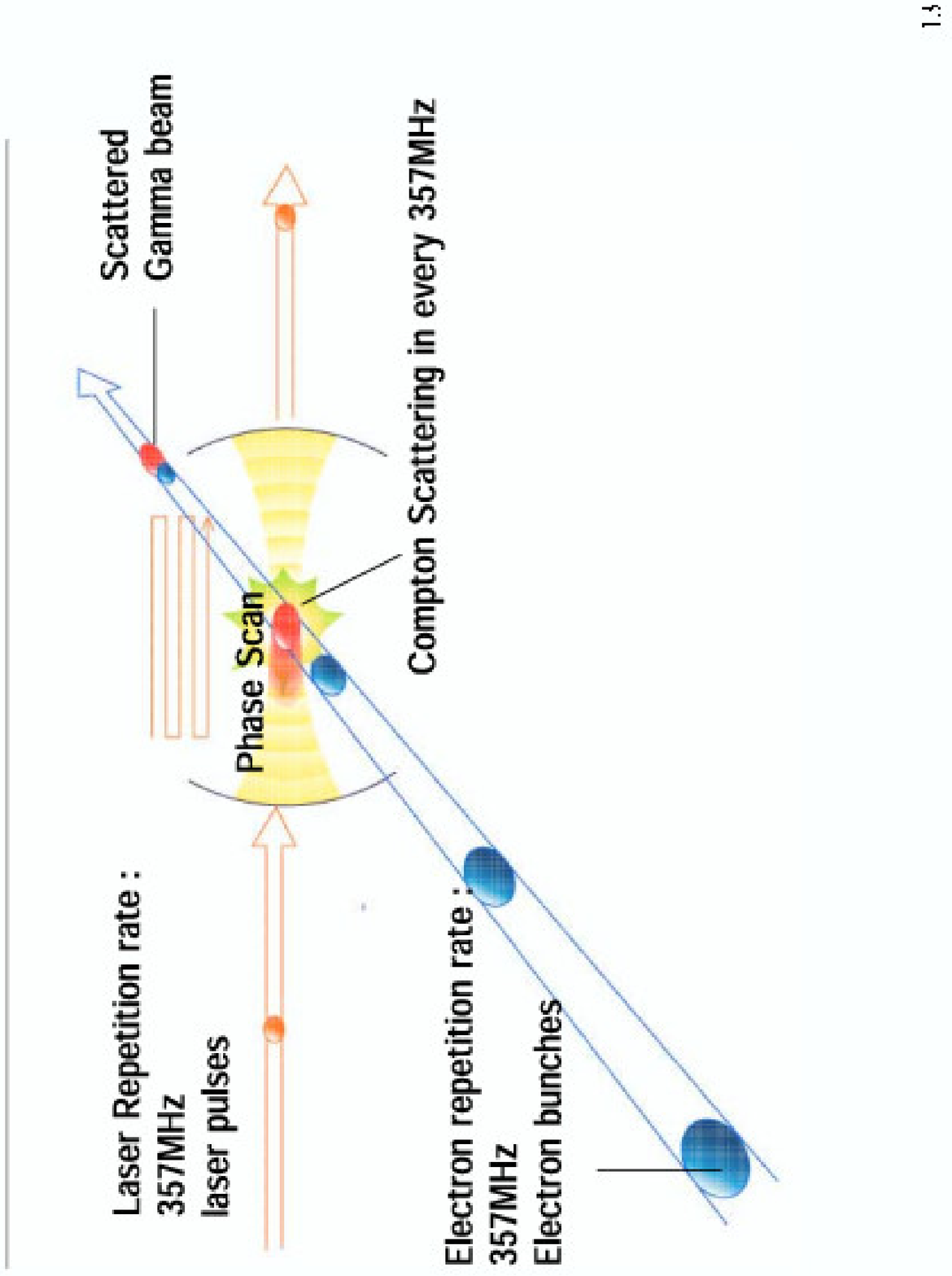}
\caption{\label{figlasercoll}Conceptual view of the  
Compton scattering experiment in ATF damping ring.}
\end{figure}

Our group \cite{Nomura} already demonstrated stable 
operation of the optical cavity with an enhancement 
factor of 230 using a 42-cm Fabry-Perot optical cavity 
made of a solid block of super-invar. 
Also, with this cavity we have conducted a laser-Compton scattering 
experiment at the ATF damping ring and measured a large rate 
of $\gamma$ rays consistent with the calculation for a 90-degree 
crossing angle between the laser and the  
electron beam \cite{Takezawa}. 
The conceptual setup of this experiment 
is shown in Figure \ref{figlasercoll}.

\begin{figure}[hbtp]
\centering
\includegraphics*[width=0.9\linewidth,clip]{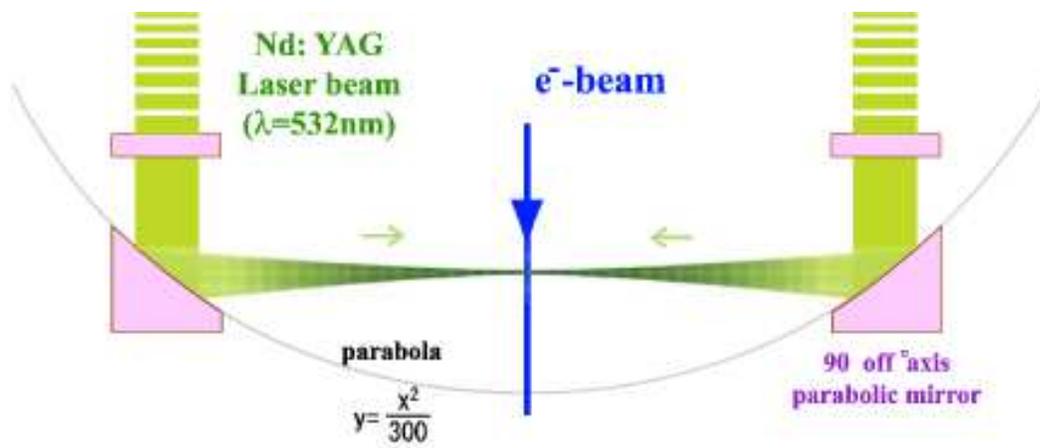}
\caption{\label{figComptonC}Conceptual view of the
ATF Compton chamber for the hybrid optical cavity.}
\end{figure}

In parallel, a Compton scattering experiment dedicated to 
polarised positron generation has also been done at the ATF,
in the extraction line. During the middle stage of this 
experiment in 2002, we used a short-focus Compton collision 
chamber consisting of two off-axis 
parabolic reflective mirrors with a 5-mm diameter hole for 
the passage of the electron beam \cite{Izumi}. 
For the next stage, we developed a hybrid optical cavity, 
in order to both produce a 
small focal spot size and to increase the enhancement factor.
The hybrid cavity 
consists of two high reflective mirrors (99.9\%) and two off-axis 
parabolic reflective mirrors (also 99.9\% reflectivity), 
shown in Figure \ref{figComptonC}. 
The mirrors are mounted on a cylindrical super-invar 
support equipped with numerous 
piezoelectric actuators to control the position of 
the four mirrors and to maintain the resonant condition 
of the optical cavity. 
Figure \ref{figComptonC} indicates three 
cylindrical arms with hole on the left 
and right for the passage of the laser pulse. 
The middle arm contains another hole for the
electron beam.

\begin{figure}[hbtp]
\centering
\includegraphics*[width=0.9\linewidth,clip]{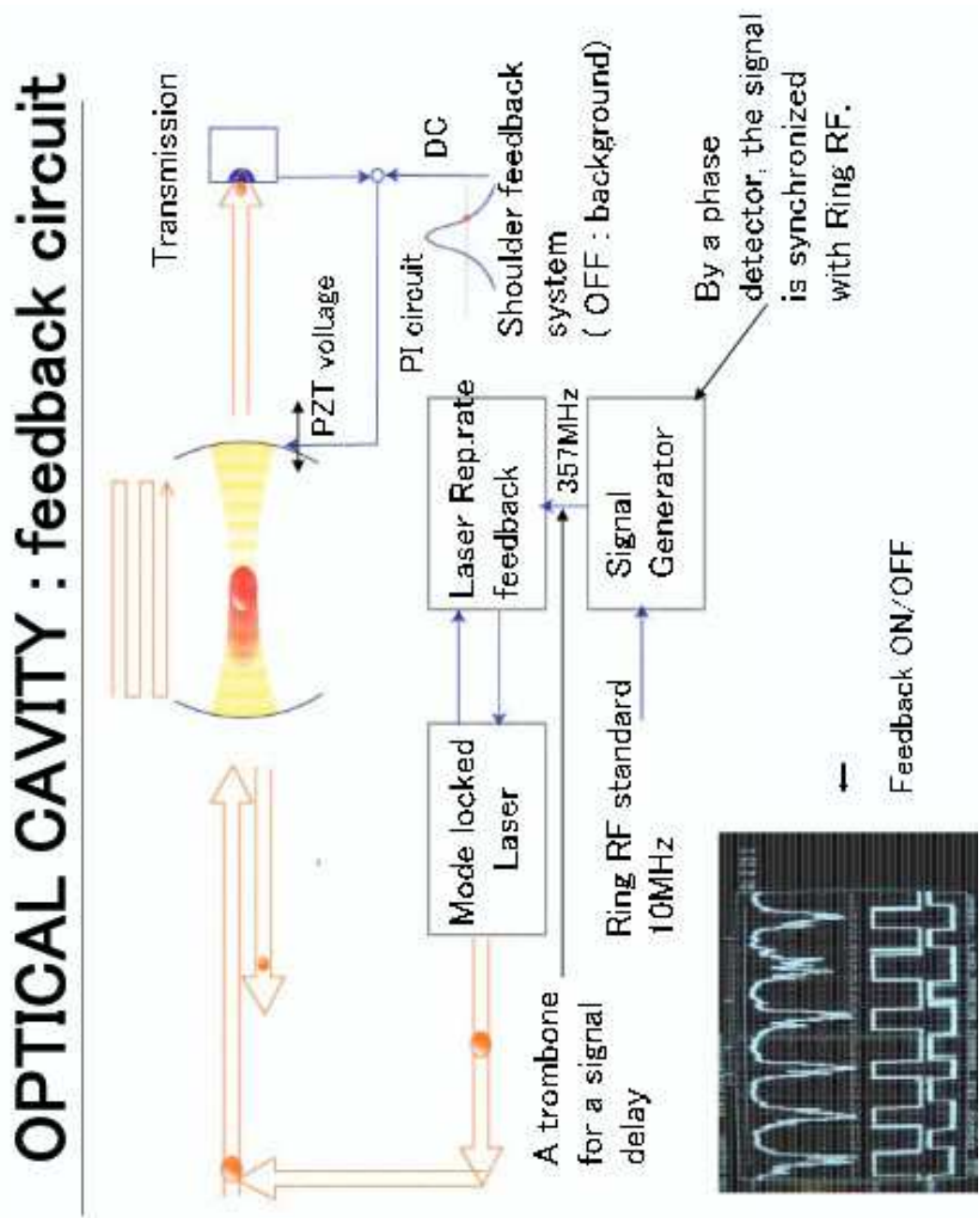}
\caption{\label{figfeedback}
Feedback system for the 
optical cavity and collision timing control
at the ATF.}
\end{figure}

Figure \ref{figfeedback} shows a schematic diagram of the 
feedback circuit employed to keep the optical cavity 
on resonance and to control the collision 
timing, which was used for a first experiment 
in 2004. This experiment at the ATF damping ring 
demonstrated a very precise collision control 
between a 10-ps laser pulse and the electron bunch.
The same techniques are also applicable for the
Compton collision chamber design of the proposed 
ILC positron-generation scheme, with the exception of 
the additional control needed for the distance 
between two adjacent Compton collision chambers. 
This inter-chamber distance should equal a large 
integer multiple of the laser wave length, in order 
to achieve a high enhancement factor and 
to maintain the collision timing of the 30 cavities. 
The position of each Compton collision chamber can be 
controlled individually, with vacuum bellows between 
neighbouring chambers. 
We are aware that an optical cavity with an 
enhancement factor of 10,000 is under development, 
but nevertheless we assume only a factor 100 
enhancement in order to arrive at a conservative design, 
given that the serial connection of 30 optical cavities 
introduces tighter tolerances, which will be estimated 
after the detailed chamber design is completed. 

\begin{figure}[hbtp]
\centering
\includegraphics*[width=0.7\linewidth,clip]{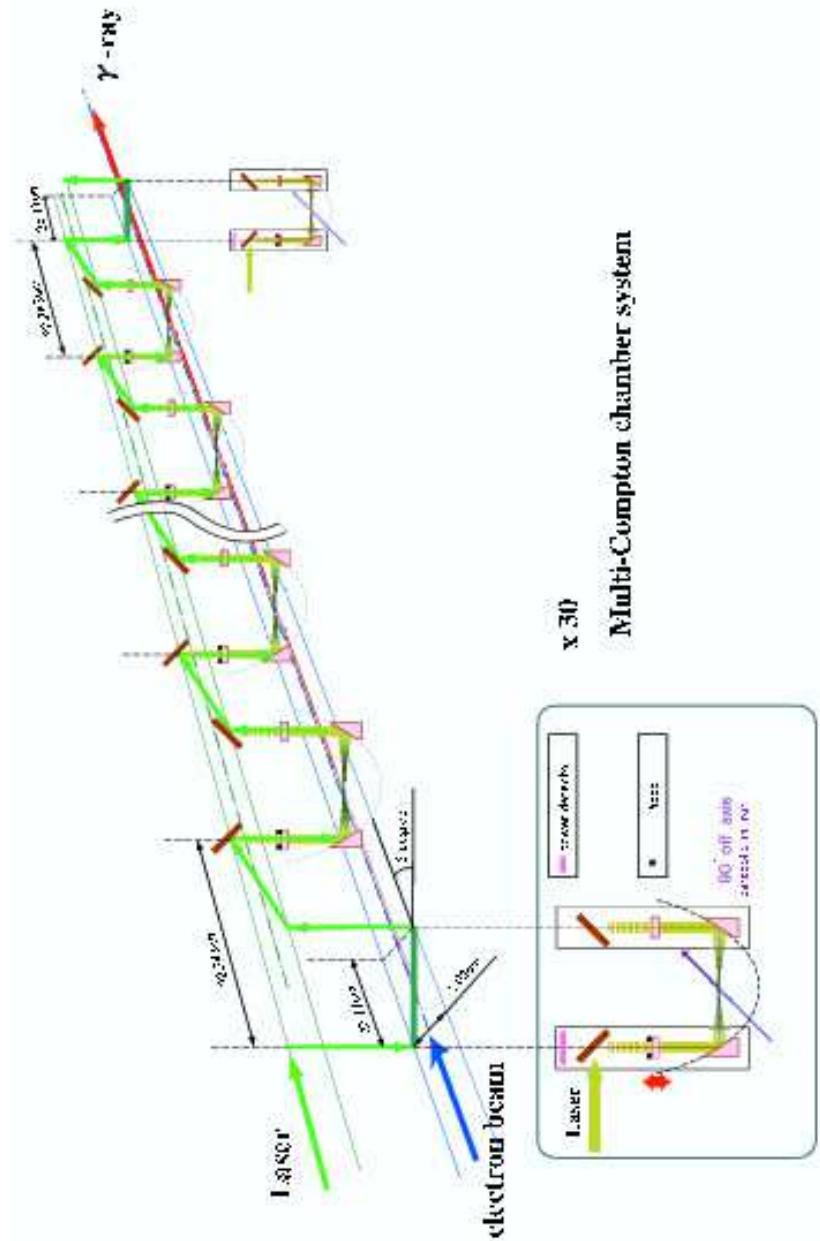}
\caption{\label{figComptonCD} Conceptual view of 
the proposed Compton collision chamber design.}
\end{figure}

The conceptual design shown in Figure \ref{figComptonCD} 
is our present proposal for the Compton collision chamber, 
which occupies about 30 m length in the straight section 
of the Compton ring. We use a FODO lattice to keep
the electron beam focused at the Compton ring IP's.

\clearpage
\sect{Injector Linac}
The injector linac accelerates the positron beam generated by the
Compton system. One pulse contains $\rm 280\times 2 \times 50 = 28000$
or $\rm 280\times 100 = 28000$ bunches. The bunch spacing is
3.08ns giving $\rm 28000\times 0.000308 = 86.2 \mu s$ pulse length.
This pulse is repeated every 10ms, with 100Hz.

If we follow the terminology of TESLA TDR, the injector linac consists
from the capture section, PPA(Pre Positron Accelerator), and MPA
(Main Positron Accelerator). 

For the capture section and PPA, it is hard to use any cold structure
due to the heavy radiation loss. On the other hand, for MPA, both
technologies, warm and cold, are possible because of the following
reasons;

\begin{itemize}
\item Average current in a pulse is 10mA. The beam loading does not
      matter for the warm linac. It does not matter for the cold linac
      too because it is identical to the main linac. Since I assume
      that the main linac will be working well, there is no doubt to
      use the same cavity for the positron acceleration.
\item Pulse length is $\rm 86.2 \mu s$. It might be critical for the
      warm linac, but the L-band warm cavity designed for the original
      TESLA positron capture section, is operated in 0.95ms duration
      with 14.5 MV/m field. $\rm 86.2 \mu s $ pulse operation
      must be much easier than 0.95ms.
\item Repetition is 100 Hz. For the warm linac, the heat load might be
      critical, but it can be cleared by assuming the same technology
      in TESLA TDR again. There is no difficulty for the cold linac
      too because the inter pulse period, 10ms, is even to fill RF
      power into the cavity whose filling time is $\rm 500\mu s$. All
      we have to do is pay a small extra cost for the cooling.
\end{itemize}

In short, MPA based on the warm and cold cavities is possible with the
conventional technologies assumed in the ILC project. In the following
sections, we will explain PPA based on the warm cavity and MPA with
the warm and cold cavities.

\subsection{Positron Pre-Accelerator}
The Positron Pre-Accelerator, PPA has a role to capture the generated
positron. It also defines the acceptance for the positron which should
be consistent to the DR acceptance, dynamic aperture. 

At this moment, the dynamic aperture is a current issue which will be
discussed in the 2nd ILC WS in Snowmass, US, but let us assume
$\varepsilon_{x,y}\sim 0.01 m.rad$ as a reference. By taking this
standard number among the ILC researchers, most of issues related to
PPA should be identical to those in other methods, undulator and
conventional\cite{PSW}.

In the original TESLA TDR, the PPA is a standing-wave normal-conducting L-band
linac\cite{TESLA}.  The front end of the PPA consists of acceleration cavities
embedded in a focusing solenoid. The first two cavities have a high
accelerating gradient ($\rm E_{acc} = 14.5MV/m$) and the others have moderate
gradients ($\rm E_{acc} = 8.5MV/m$). Each cavity is powered by one 10MW
klystron.

The only difference of the PPA in the Compton method is the repetition
and the pulse duration. The averaged current in a pulse is designed to
be identical.

In other methods, the PPA is driven in 5Hz with almost 1ms duration.
These numbers are 100Hz with 100ms interval with almost $\rm 100\mu s$
duration in this method. The heat load to the cavity is estimated to
be exactly same as that in the other method since the ratio of the
heat load to that of the other methods, $C_{hl}$ is accounted as
\begin{equation}
C_{hl}=\frac{100}{5}\times \frac{100}{1000}\times \frac{1}{2} = 1.0,
\end{equation}
where the first fraction comes from the repetition rate, the second term
means the ratio of the pulse duration, and the last term means the duty
cycle, i.e. the cavity is driven in half of every 100ms.

>From this simple estimation, we do not have to pay any extra cost to
drive the PPA with this high repetition and short duration mode.  We do
not need any modification on the PPA.

The gradients in the first and second cavities of the PPA are limited by
RF power and heat load restrictions. Three dimensional thermal stress
analysis indicates that stable and reliable cavity operation with a heat
load of about 30 kW/m is possible, corresponding to an accelerating
gradient $\rm E_{acc} = 14.5MV/m$ with the long RF pulse ($\rm 950\mu s$
flat-top) and repetition rate of 5Hz\cite{KFloettmann}.  The heat load
of the cavity in the same gradient with the high repetition and short
duration is identical, this statement is therefore also true for our
case.

The rest of the PPA consists of five cavities with moderate gradient
(8.5MV/m) as same as that in TESLA TDR\cite{TESLA}. Each cavity has two
accelerating sections. The transverse focusing is accomplished using
quadrupole triplets, placed between the sections.  Each cavity is
powered by one 10MW klystron. 

At the exit of the PPA section, 287 MeV bunched positron beam is
obtained. 

\subsection{MPA based on the warm technology}
In MPA, the same technology as the latter part of the PPA section can be
used to accelerate the beam up to 5 GeV.

One accelerating module consists from two accelerating sections and
quadrupole triplets\cite{KFloettmann}. Total length of the module is
4.3m. The energy gain per one module is 34.6 MeV. To accelerate the
positron beam up to 5 GeV, 144 modules are required including 5\% of the
margin. The total length of the linac becomes 690m including 0.5m for
the quadrupole triplets inserted between the modules. One module is
driven by one 10MW klystron, so that totally 144 klystron is needed. The
parameters of MPA based on the L-band warm structure, are
summarised in Table \ref{warm-table}.
\begin{table}
\centering
\caption[]{Parameters of the injector linac based on the L-band warm
structure.\label{warm-table}}
 \begin{tabular}{@{\vrule width 1pt}c|r|l@{\ \vrule width
  1pt}}\hline\hline
Item & number & unit \\\hline
Field gradient & 8.5 & MV/m\\
Energy gain per module & 34.6 & MeV\\
Number of module & 144 & unit\\
Number of klystron & 144 & unit\\
Total length & 620 & m \\\hline\hline
 \end{tabular}
\end{table}

\subsection{MPA based on the cold technology}
If we employ the cold technology for MPA, we can use the system almost
identical to that of the main linac. The bunch spacing is 3.08 ns which
is 100 times smaller than that in the main linac, but the bunch
intensity is only 0.03nC giving the exactly same average current,
10mA. Because of this fact, we can use the same cavity and coupler
including the coupling coefficient.

If the cavity is operated in 100 Hz repetition with $\rm 100\mu s$
duration, the heat load generated when the beam is on, can be identical
to that in the main linac because of the same average current and
effective pulse duration, $100\mu s \times 10 = 1 ms$.  The heat load
coming from the RF filling and decay time, however, contribute and
dominates the total heat load.

If the heat load from the RF filling and decay period corresponds to
that of $\rm 500\mu s$ of the beam period, the multiplication factor of the
total heat load of the Compton operation mode against to that of 1ms operation,
$C_{RF}$ is
\begin{equation}
C_{RF} = \frac{500\times 10 + 100\times 10}{1000 + 500}=4.0
\end{equation}
We need therefore 4 times larger cooling power to operate MPA in the
Compton mode. This excess shares, however, only $\rm
15/(500+500+5+5)=1.4$\% and $\rm 15/(1000+1000+5+5)=0.7$\% of the total
heat load in the whole ILC 500 and 1000 respectively.

By considering the large emittance compared to the main linac, we have
to employ a same arrangement of the cryomodule of that in TESLA TDR. To
achieve enough focusing to accelerate the positron beam which has a
large emittance, two types of the cryomodule assemblies are
designed\cite{TESLA}.

The cryomodule for MPA is implemented from the two types of the standard
ILC cryomodules. The transverse focusing is carried out by quadruples
doublets. Type 1 module consists from four accelerator cavities and four
quadrupole doublets. Type 2 module consists from 8 accelerator cavities
and 1 quadrupole doublets. The length of the cryomodules are 11.4m and
12.4 m for type 1 and type 2 respectively.

In the TESLA TDR, the field gradient in the cavity is assumed to be 25
MeV/m\cite{TESLA}. If we employ 35 MV/m which is likely to be a new
standard of the field gradient in ILC, the energy gain per the module
can be 140 and 280 MeV/module for type 1 and 2 respectively.  Assuming
this higher acceleration field than that in TESLA MPA module, we can
reduce the number of the modules to be 6 of type 1 and 15 of type 2.
Number of modules in the original TESLA TDR was 8 of type 1 and 20 of
type 2.

The total energy gain of MPA is estimated to be 5040 MeV.  Including the
energy gained by PPA, the final energy of the positron beam becomes 5290
MeV which is even enough by assuming 5\% margin.  Total length of MPA
becomes 270m. This number includes 0.5m for the inter-module bellows
section and 5m for the cooling channel.  One 10MW klystron can drive 10
cavities with 35 MV/m gradient\cite{KSaito}.  Total number of cavities
in MPA is 139 which requires 14 klystrons.  Considering a margin, let us
assume 3 klystrons for 6 type1 modules and 12 klystrons for 15 type 2
modules. The total number of klystrons is 15.  Parameters of MPA based
on the cold cavity are summarised in Table\ref{cold-table}.
\begin{table}
\centering
\caption[]{Parameters of MPA based on the cold technology. \label{cold-table}}
 \begin{tabular}{@{\vrule width 1pt}c|r|l@{\ \vrule width 1pt}}\hline\hline
 Item  & Number & unit\\\hline
Number of type 1 module & 6 & unit\\
Module length & 11.4 & m\\
Energy gain/module & 140 & MeV\\
Number of quadrupole doublets & 4 & unit\\
Number of cavities & 4 & unit\\\hline
Number of type 2 module & 15 & unit\\
Module length & 12.4 & m\\
Energy gain/module & 280 & MeV\\
Number of quadrupole doublets & 1 & unit\\
Number of cavities & 7 & unit\\\hline
Final energy & 5.29 & GeV\\
Total length & 270 & m \\
Total number of cryomodules & 21 & unit \\
Total number of klystrons   & 14 & unit \\\hline\hline
 \end{tabular}
\end{table} 


\clearpage
\sect{Total System}





\begin{figure*}[hbtp]
\includegraphics*[height=\linewidth,angle=90]{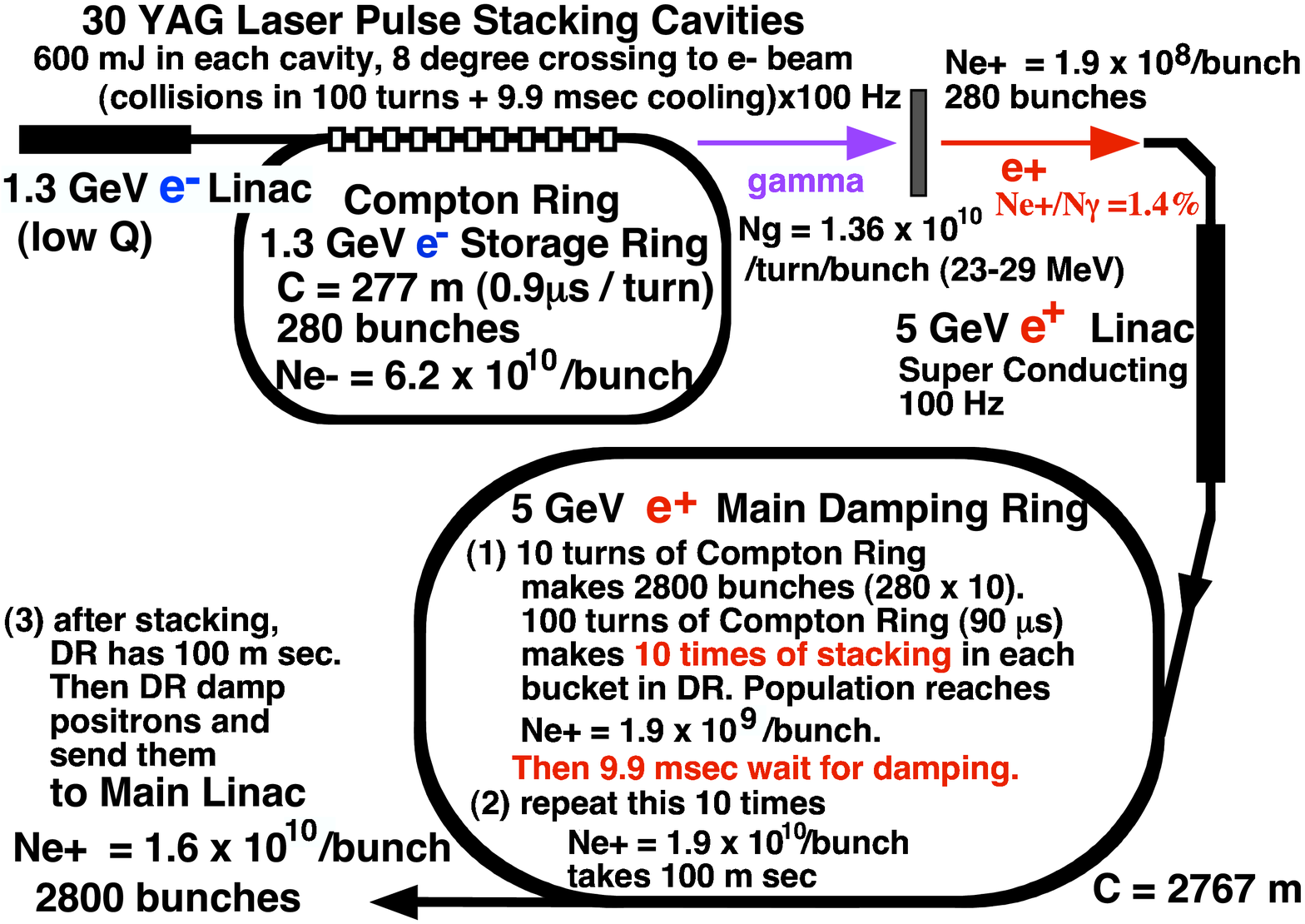}
\caption{\label{figYAG}Schematic diagram of the total system in the case 
  of YAG laser.}
\end{figure*}

\begin{figure*}[hbtp]
\includegraphics*[height=\linewidth,angle=90]{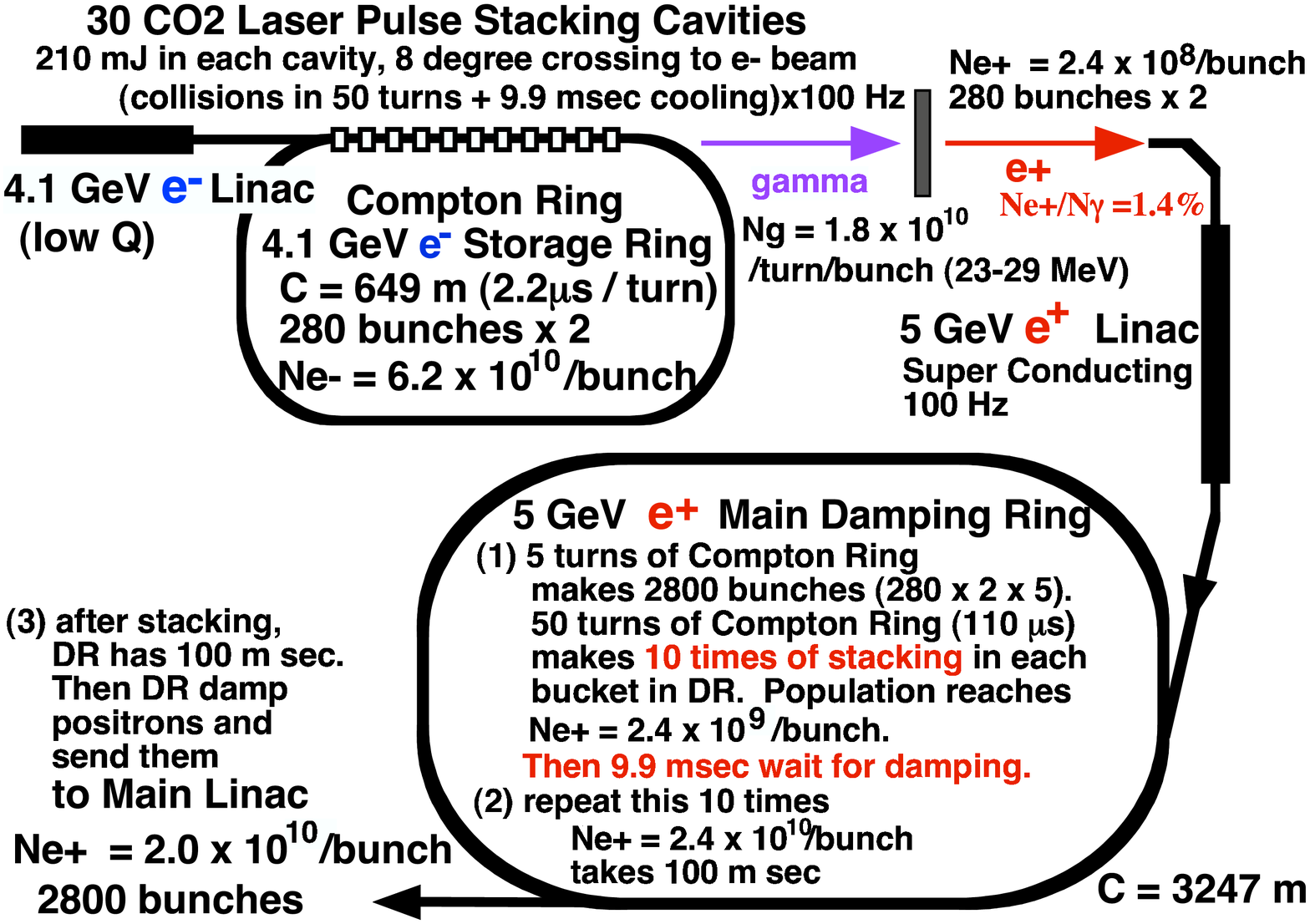}
\caption{\label{figCO2}Schematic diagram of the total system in the case 
  of CO2 laser.}
\end{figure*}

Figures \ref{figYAG} and Figure \ref{figCO2} illustrate 
the total system for two versions of our proposal. 

The first represents the case of YAG laser Compton scattering. It assumes that
one train (280 bunches) of the electron beam is stably circulating in the
Compton ring except for the period of the Compton scattering (100Hz, about
$100\,\mu$s pulse width, duty factor $50\%$ ).  Electron losses are not
problem, but bunch lengthening is a concern since the $\gamma$-ray yield
depends on the bunch length due to crossing angle.  From our beam tracking
simulation for the Compton ring, we infer that the beam loss is negligible and
that the electron bunch recovers in about 9.9 ms after a 100-turn burst
collision.  A more detailed design study and optimisation of the Compton ring
will necessary prior to construction.  This $\gamma$-ray generation scheme
requires a 100-Hz 5-GeV injector linac after the conversion target and 10
times successive beam stacking into each bunch of the main damping ring, which
is repeated for 10 linac pulses. If we assume appropriate beam parameters, the
beam tracking simulation for the damping ring shows that 10 times successive
beam stacking is possible. The conceptual design of the Compton collision
chamber and the laser system are described in Sections 5 and 6.

A preliminary Compton collision experiment has been performed 
using pulsed YAG laser (10ps FWHM) and a simple Fabry-Perot 
optical cavity with a 90-degrees crossing angle. 
The state of art was already demonstrated with about a factor 
100 enhancement of the optical cavity, except for technologies 
and issues related to a small crossing angle geometry and 
the more strongly focused laser beam size. The experimental 
test of a double Compton-chamber system is necessary and 
under planning at the ATF damping ring. 

An alternative system uses a CO2 laser. The principle of this 
scheme is analogous the case of the YAG laser. One difference is that 
the beam energy in the CO2-laser Compton ring is 4.1 GeV 
instead of 1.3 GeV for the YAG laser. 
The CO2 laser uses a 50-turns burst collision, instead
of a 100-turn collision, The 50-turns collision 
generates sufficient $\gamma$-rays 
in the Compton ring, which has a larger circumference
and stores a beam of two trains (2x280 bunches). 

The average positron population of the injected bunches 
is about $2 \times 10^8$ for either laser, and this intensity 
changes dynamically by a factor of about 4 during the 
Compton collisions. Since the beam loading is very small, 
we do not expect any problem in the 5-GeV injector linac 
and can control the stacking beam intensity in the main 
damping ring. The most important outstanding technical 
issues are to develop the laser system and the Compton 
collision chamber. 

The polarised-positron generation scheme which
we propose is very flexible, and of moderate size.
It provides a fully independent system which means 
that we can perform the ILC beam commissioning 
at full beam power without the need of a 150-GeV 
electron beam.  The design of the Compton ring, 
the Compton collision chamber, and the laser system 
will be optimised with respect to tolerances.

\clearpage
\appendix
\sect{Cavity R\&D in Orsay}
\label{sec:rdorsay}
\subsection{Introduction}
The envisaged high finesse of around 30000 has already been reached for
cavities filled with a continuous Nd:YAG laser beam in polarimeters used at
CEBAF \cite{nico,cebaf-pap,baylac} and DESY \cite{habil}.
Also some cavities are already working in the pulsed regime.
At SLAC a cavity with 30\,ps pulses reached a finesse of 12000 \cite{slac}.
At KEK a pulsed cavity is built for a laser wire application \cite{Nomura}.

\subsection{Description of present project}

\begin{figure}[htbp]
  \begin{center}
    \leavevmode
    \includegraphics[width=\linewidth]{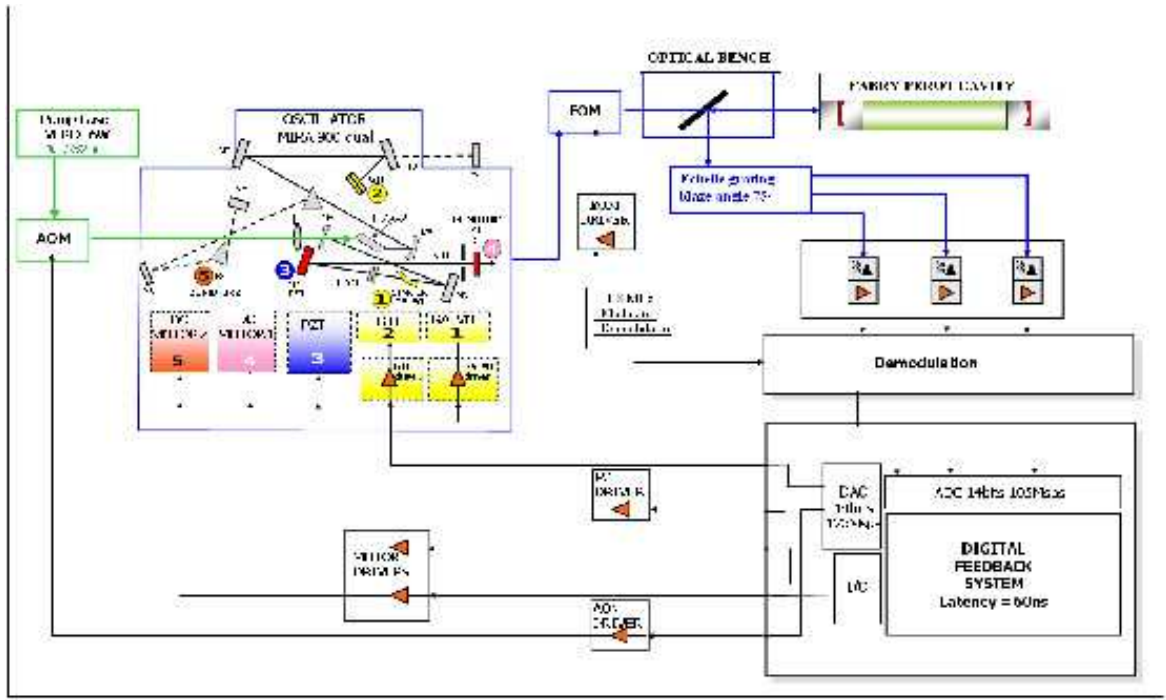}
   \caption{ Schematic view of the feedback system. See text for a description
     of each component. In the MIRA oscillator, mirrors M6, M7 and M9 (M10
     being removed) and the two prisms P1 and P2 are used in the 100fs regime.
     In the 1ps, the prism P1 is moved away from the optical path and M10 is
     inserted. The Lyot filter \cite{lyot} and the output coupler M1/OC are
     also changed.}
    \label{fig-orsay}
  \end{center}
\end{figure}

\begin{figure}[htbp]
  \begin{center}
    \leavevmode
    \includegraphics[width=\linewidth]{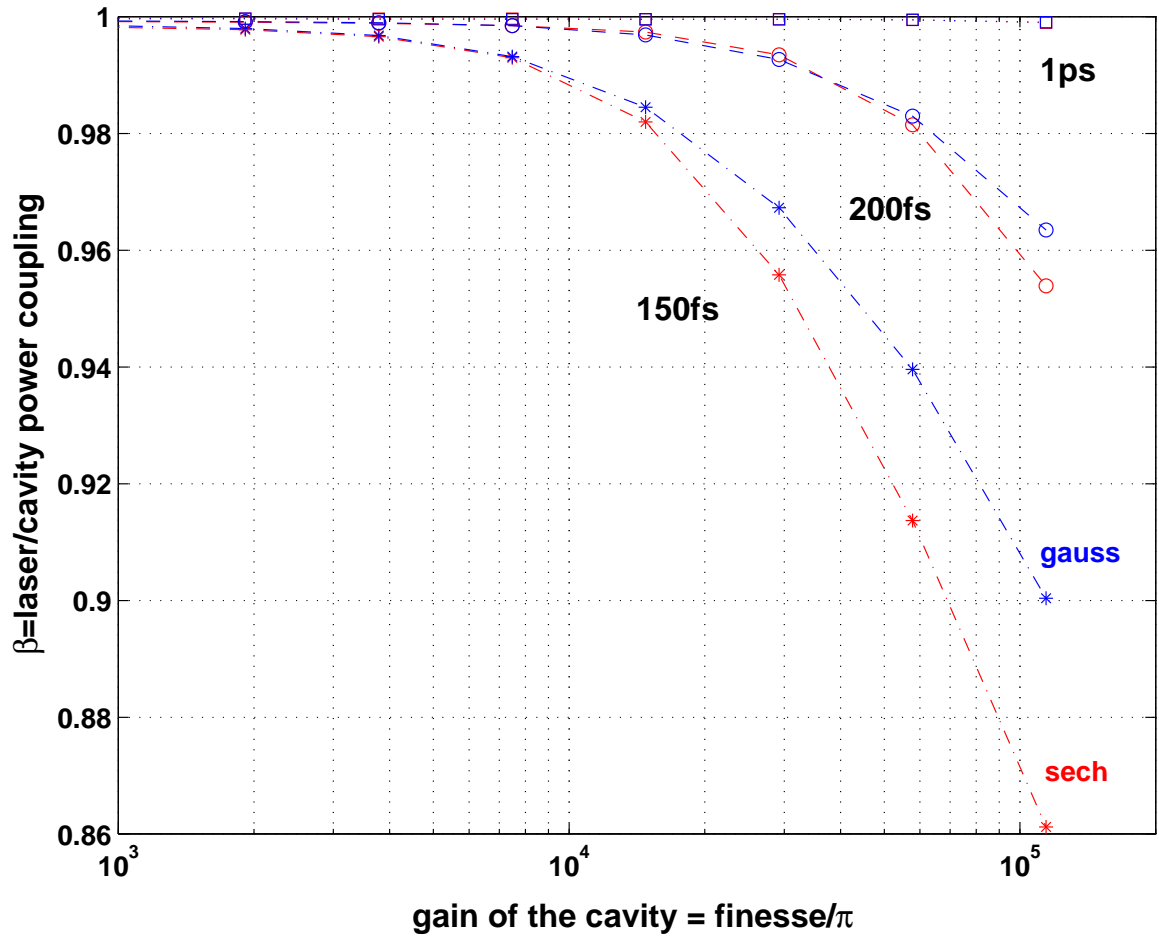}
   \caption{Laser-beam/cavity-beam coupling as a
 function of the gain of a 2m long Fabry-Perot cavity.
 Only the chromatic dispersion effects in the cavity
 mirror coatings are taken into account in the coupling calculation. To pulse shapes
 are considered: a Gaussian and a soliton (denoted by sech on the figure).}
    \label{fig-sech}
  \end{center}
\end{figure}

Our R\&D is an investigation on whether a high finesse Fabry-Perot resonator
\cite{kogelnick} can be used to `amplify', with an energy gain ranging from
$10^4$ to $10^5$, a passive mode-locked laser beam \cite{Krautz}. This would
allow a drastic gain in the Compton cross section once that the cavity is
located around the electron beam.  Two pulse laser time widths will be tried:
full width at half maximum (FWHM) of 100fs and 1ps. 

 In a first step, a
confocal (mechanically stable) two mirrors cavity will be considered. High
quality mirrors will be mounted in order to reach cavity finesses of 30000 and
300000 ({\it i.e.} power gain of 10000 and 100000).  In a second step, a more
complex resonator will be considered to reduce the laser beam waist inside the
cavity : a concentric cavity (mechanically unstable) made of two mirrors or a
four-mirror cavity. This second step is still under study at the present time.
Both steps require a special mechanical design of highly stable cavity mirror
mounts and a great care on the reduction of the environment noise.

In principle, the smaller the pulse time FWHM the larger the laser frequency
spectrum. One then expects that chromatic dispersion induced by the pulse
propagation in the multilayer mirror coatings
\cite{yecoupling1,yecoupling2,peterson} and the finite frequency bandwidth of
these coatings \cite{yemirror} will reduce the coherent coupling of the
incident laser pulses to the pulse circulating inside the cavity. We performed
these calculations for soliton and Gaussian like pulses.
 The cavity mirror reflection coefficients were calculated numerically 
 using standard multilayers formula \cite{pochi} including the values
 of the optical indices as given by our mirror coating
 manufacturer \cite{lma}. Fig. \ref{fig-sech} shows the effective cavity gain $\beta$
 as a function
 of the gain of a 2m long Fabry-Perot
 cavity ({\it i.e.} the number of double layers constituting the 
 cavity mirror coating). From this figure one sees that this effect
is negligible for 1ps FWHM whereas the coupling of the incident power is
reduced by $\approx 14\%$ for 150fs FWHM. Pulses with time FWHM as small as 150fs
 can therefore be envisaged to be efficiently `amplified' in a very high finesse
 Fabry-Perot cavity. Note that passive mode locked laser pulses 
 have a time shape closed to that of a soliton. 

The experimental scheme of the first step of our R\&D is shown in Fig.
\ref{fig-orsay}.  A commercial passively mode locked laser, Coherent's MIRA
Ti:sa oscillator with 76MHz pulse repetition rate, pumped by a green laser
beam, Coherent's 6W VERDI, is sent into a Fabry-Perot cavity.  The laser is
locked to the cavity by means of the Pound-Dever-Hall technique \cite{pound}
adapted to the pulsed laser beam regime
\cite{feed-pulse,ye,lautre,ye-stabilisation,book-fscomb}.  Following our
estimation both the laser pulse repetition rate $f_{rep}$ and the phase shift
between the carrier and the envelope of the electric field $\varphi_{ce}$
\cite{udem} must be locked to the Fabry-Perot cavity in the two pulse length
regimes (1ps and 100fs).  To perform this locking, at least two error signals
are necessary. To build the error signals, the laser beam is phase modulated
by an Electro-Optic-Modulator (EOM) and the signal reflected by the cavity is
sent to a diffraction echelle grating. The intensity is then read at different
positions, that is at different values of the wavelength. After demodulation,
these signals are sampled and analysed by a digital feedback system (Lyretec
XXX card). Correction signals are built and sent to various actuators
according to their frequency bandwidths and functionalities:
\begin{itemize}
\item an Acousto-Optic-Modulator (AOM), used as an amplitude modulator for the
  pump laser beam to control the fast variations of both $f_{rep}$ and
  $\varphi_{ce}$ \cite{ye-calib};
\item a fast piezoelectric transducer glued on one of the oscillator mirror to
  control the fast variations of $f_{rep}$;
\item a galvometer on which is located the mode locking starter (= two
  mirrors) to follow the slow drifts of $f_{rep}$;
\item a translation stage on which is located the oscillator output coupler to
  adjust roughly the oscillator length to the Fabry-Perot cavity length;
\item a piezoelectric transducer acting on the width between the platelets of
  the Gire-Tournois (GTI) interferometer \cite{gire}. With this actuator, one
  is able to follow the slow drifts of $\varphi_{ce}$ in the 1ps version of
  the MIRA ($f_{rep}$ being also affected).
\item a translation stage on which is located one of the prism in order to
  control the slow variations of $\varphi_{ce}$ in the 100fs version of the
  MIRA ($f_{rep}$ being also affected).
\end{itemize}

The feedback strategy is not, and cannot \cite{ye-calib}, be defined {\it a
  priori}, {\it i.e.}  one must calibrate the pump and oscillator lasers to
define which actuators must be used in conjunction with a given error signal.
This calibration depends on our lasers and on the environment noises. To
calibrate the actuator responses, we shall follow the methods of ref.
\cite{ye-calib,witte}.  To characterise the phase and amplitude laser noises
\cite{noise-theo}, we shall use the setups of refs.
\cite{noise-measure1,noise-measure2}. With our 'long' pulses, an
autocorrelator \cite{xu-autoco} will be used to observe the variations of
$\varphi_{ce}$ during the calibration experiments.  As for $f_{rep}$, a
spectrum analyser and a large band oscilloscope will be used.  Finally, the
beam pointing stability of the pump laser beam will be determined as described
in \cite{witte}.

As previously mentioned, the mechanical stability will play a fundamental role
in the second phase, where we envisage to pass from a confocal configuration
to a concentric one.  This will allow us to obtain a much smaller waist size
in the Fabry Perot resonator,thus increasing the photon density in the
interaction point.

The main problem is that
the concentric configuration is highly mechanically unstable. 

As an example we can say that in our 2m long cavity a waist of around 130
microns can be obtained with a 1 cm separation between the two centres of
curvature of the two mirrors. This will imply a maximum range of stability
that corresponds to less than 10 microradian in tilt (for this tilt the
optical axis is out of the cavity).  Reducing the waist size will require more
stringent constraints.

Therefore much attention will be paid to the environmental noise suppression
and for the optical table passive stabilisation. The two mirrors of the cavity
will be mounted on separate holders allowing an independent distance
regulation.

Furthermore, to approach this problem we have studied a new mirror holder
system (see fig 2) that will be tested on the confocal cavity as a proof of
principle.  The principle of this holder is based on a monoblock cardan joint
system that allows both rotations around the geometrical centre of the mirror.
Thanks to its specificity this system is based on simple flexions, thus
avoiding backlask and friction. The external tilt adjustment is assured by a
wedge system.

\begin{figure}[htbp]
\centering
\includegraphics[width=\linewidth,clip]{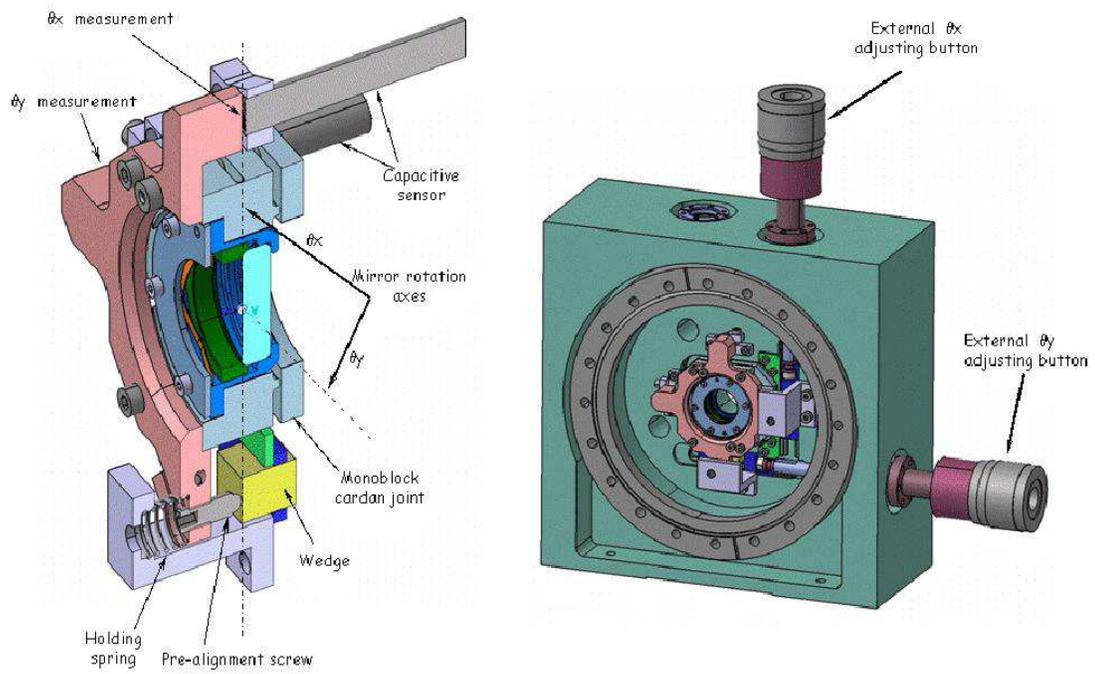}
\caption{\label{f:fig2}
Mirror holder prototype}
\end{figure}

This avoids the stacking of different elements (therefore adding vibration
sources), minimises the distance between the regulation point and the support,
and the mechanical coupling with the environment.  A very good sensitivity for
the displacement is ensured by the high demultiplication factor that gives a
precision of $\approx$ 0.5 microradian.  A capacitive sensor is installed to
measure the mechanical stability and the tilt and to eventually provide the
signal for an active feedback. This is installed on the support and reads the
distance between the support itself and the mirror holder. The resolution is
20 nm at 1 kHz of acquisition rate.


\clearpage
\section*{Acknowledgements}
The present research has been financially supported by a Grant-In-Aid
for Scientific Research for Creative Scientific Research of JSPS (KAKENHI
17GS0210) and by the European Community-Research Infrastructure
Activity under the FP6 ``Structuring the European Research Area'' programme
(CARE, contract number RII3-CT-2003-506).

\end{document}